\documentclass[a4paper,11pt]{article}

\pdfoutput=1
\usepackage{multirow}
\usepackage{jcappub} 
\usepackage{float}
\usepackage{mathtools}
\usepackage{comment}

\usepackage[T1]{fontenc}
\usepackage{subfig}
\usepackage{booktabs}
\title{\boldmath Fisher matrix for the angular power spectrum of multi-tracer galaxy surveys}

\author[a]{L. Raul Abramo}
\author[a]{, Jo\~ao Vitor Dinarte Ferri}
\author[a]{, Ian Lucas Tashiro}
\author[b,c,d]{and Arthur Loureiro}

\affiliation[a]{\small Departamento de F\'{\i}sica Matem\'atica, Instituto de F\'{\i}sica, Universidade de S\~ao Paulo,\\ R.  do  Mat\~ao  1371,  05508-090,  S\~ao Paulo, SP, Brazil}

\affiliation[b]{\small Department of Physics and Astronomy, University College London, Gower Street, London WC1E 6BT, UK}

\affiliation[c]{\small Institute for Astronomy, University of Edinburgh, Royal Observatory, Blackford Hill, Edinburgh EH9 3HJ, UK}

\affiliation[d]{\small Astrophysics Group, Blackett Laboratory, Imperial College London, London SW7 2AZ, UK}

\emailAdd{abramo@if.usp.br}
\emailAdd{joao.vitor.ferri@usp.br}
\emailAdd{iantashiro@usp.br}
\emailAdd{arthur.loureiro@ed.ac.uk}

\abstract{
Redshift evolution and  peculiar velocities break the isotropy of cosmological surveys with respect to the directions parallel and transverse to the line of sight, limiting the accuracy of the Fourier representation to small areas and redshift ranges.
In contrast to the Fourier space power spectrum, the full information about the two-point function of tracers of large-scale structure is encapsulated in the redshift-dependent angular power spectrum  $C_\ell^{ij} (z_i,z_j)$ for the tracer species $i$ and $j$ at the redshift slices $z_i$ and $z_j$, expressed in harmonic space.
In this paper we derive semi-analytical expressions for the multi-tracer Fisher matrix of angular power spectra, in real and in redshift space, which are exact in the linear regime of structure formation.
Our expressions can be used to forecast the constraining power of galaxy surveys with many tracers and a large number of redshift slices, for which the derivation of the Fisher matrix from numerically evaluated covariance matrices may not be feasible or practical.
}


\begin{document}

\maketitle
\flushbottom

\section{Introduction}
\label{sec:introduction}

Small-area, shallow astrophysical surveys are increasingly giving way to large-area, deep surveys that can map the distribution of galaxies and other tracers of large-scale structure over large fractions of our past light-cone.
This process, which started with the Sloan Digital Sky Survey \cite{SDSS}, is now being driven by surveys such as DESI \cite{DESI}, PFS \cite{PFS}, Euclid \cite{Euclid}, the Vera Rubin Observatory Legacy Survey of Space and Time \cite{LSST}, J-PAS \cite{JPAS}, and SKA \cite{SKA-2020PASA}.
These surveys will go beyond just measuring the scale of baryon acoustic oscillations, providing exciting new tests of modified gravity \cite{ModGrav2012,ModGrav2016}, primordial non-Gaussianities \cite{Dalal2008}, relativistic effects \cite{Yoo2010PhRvD..82h3508Y,Relativistic2011PhRvD..84d3516C,BonvinDurrer2011,Yoo2012PhRvD}, as well as disentangling the cosmological mechanisms that operate at ultra-large scales \cite{DisentanglingBruni,Disentangling2017,Caroline2021MNRAS}.

It is therefore critical that we understand to what extent future surveys will be able to capture the information contained on these largest observable scales, both in terms of the vast redshift ranges that are now accessible to our instruments, but also in terms of large angular separations that will be covered by those surveys.
This means, in particular, that a traditional approach based on the Fourier-space power spectrum cannot possibly provide an accurate description: in its simplest form, it assumes that the sky is flat, which is equivalent to assuming that the survey is inscribed by a Cartesian box placed at a large distance from the observer.  
More sophisticated approximation schemes that try to take into account the curvature of the sky are able to partially offset this problem \cite{Yamamoto2006,Bianchi2015,Scoccimarro2015}, but they also start to fail at intermediate scales.
We are then led almost inevitably to a spherical harmonic description of the matter distribution in redshift space \cite{Peebles1973,FSL1994MNRAS,
HeavensTaylor1995,1996MNRAS.278...73H,2008GeoJI.174..774D}, which should retain information from large angular scales and all redshift ranges, as well as taking into account the many different tracers of large-scale structure: 
galaxies \cite{Scharf1992MNRAS.256..229S,Huterer2001ApJ...555..547H,Blake2004MNRAS.351..923B,Padmanabhan2007MNRAS.378..852P,Blake2007,Ho2012ApJ...761...14H,Asorey2012MNRAS.427.1891A,Andres2018MNRAS.476.1050B,Xavier2019JCAP...08..037X,Hugo2019MNRAS.487.3870C,Loureiro2019MNRAS.485..326L,Felipe2021MNRAS.505.5714A}, 
quasars \cite{Leistedt2013MNRAS.435.1857L},  as well as radio and
HI intensity mapping \cite{Ferramacho2014MNRAS.442.2511F,Fonseca2015ApJ...812L..22F}.
The description of weak lensing in terms of a spherical harmonic decomposition and angular power spectra has also become increasingly popular 
\cite{Huterer2002PhRvD..65f3001H,Hikage2011MNRAS.412...65H,Loureiro2021arXiv211006947L,Upham2021arXiv211207341U}.

One of the observables that can encapsulate all the information of our past light-cone is the angular power spectrum $C^{ij}_\ell (x_i,x_j)$, which is the harmonic component ($\ell$) of the clustering between the tracer $i$ at the redshift slice $z_i$ at radius $x_i=\chi(z_i)$ and the tracer $j$ at the redshift slice $z_j$ and radius $x_j=\chi(z_j)$. 
However, given the large number of degrees of freedom in these angular power spectra, it can be very difficult to estimate the constraining power of large surveys on the basis of these observables. 
Despite recent breakthroughs in numerical implementations of the harmonic approach \cite{3DEX,LibSharp2013A&A...554A.112R,Namaster2019MNRAS.484.4127A,SFB2021}, analytic and semi-analytic tools are still needed.
In particular, spectroscopic and intensity mapping surveys allow us to split the observations into a large number of thin redshift slices, which then become increasingly correlated. 
By using thinner redshift slices we avoid smearing the maps of the tracers over the radial direction, but this creates a problem insofar as the covariance matrix for the angular power spectra grows fast with the number of slices: e.g., even for a single tracer, splitting a survey into 50 redshift slices leads to a $(1275)^2$ covariance matrix for each multipole $\ell$.
Computing such an object with sufficient accuracy may be fraught with difficulties, especially given that what we actually need is either the inverse of that covariance (i.e., the Fisher matrix) or its Cholesky decomposition. 
From a different perspective, with the increasing popularity of Bayesian Hierarchical Models for Field-Level Inference\cite{Jasche-2013MNRAS.432..894J,Alsing-2016MNRAS.455.4452A,Alsing-2017MNRAS.466.3272A,Ramanah-2019A&A...621A..69R,Lavaux-2019arXiv190906396L,Porqueres-2020A&A...642A.139P,Leclercq-2021MNRAS,Porqueres-2022MNRAS.509.3194P}, analytic expressions for the covariance matrix and its inverse have an important application as the mass matrix or step-size proxies in a Hamiltonian Monte-Carlo framework \cite{Jasche-2010MNRAS.407...29J,Betancourt-2017arXiv170102434B}, allowing for a significant increase in efficiency for these samplers.
As is the case for any Fisher analysis, we do assume Gaussianity (since that is what enables us to write the covariance of the power spectra in terms of a 4-point function), but our semi-analytical results are able to capture the full time evolution along the past light-cone.
In this paper we derive a semi-analytical expression for the multi-tracer Fisher matrix of the angular power spectra.
Our results will pave the way to forecasting accurately the constraining power of surveys on ultra-large scales, and to estimate the uncertainties in the angular power spectra.
In particular, since we do not employ the Limber approximation \cite{Limber}, we are able to obtain an analytical expression for the radial mixing matrix that determines the coupling between radial modes in the angular power spectra.
Finally, we should stress that in this paper we do not resort (at least not explicitly) to the spherical Fourier-Bessel transform of the density field \cite{HeavensTaylor1995,Abramo2010PhRvD,3DEX,SFB2021}. 

This paper is organized as follows. In Section 2 we review general expressions for the covariance and Fisher matrices, including the case when there are multiple tracers, and including their cross-correlations -- always under the assumption of Gaussianity.
In Section 3 we show how to write the {\it data covariance} (i.e., the observed angular power spectra, including shot noise) and how it is possible to invert that object -- which is the key step that allows us to compute the Fisher matrix.
Still in Section 3 we present an exact solution for a toy model, that also serves as a check for our formulae.
In Section 4 we obtain the data covariance in redshift space, and present a  semi-analytical expression for its inverse -- the Fisher matrix.
At the end of that Section we also discuss in which circumstances one would employ our semi-analytical methods, as opposed to direct numerical computations in terms of mocks or simulations.
We conclude with a discussion in Section 5.
For the sake of clarity and readability, some  technical details are left for the three appendices.

\section{Fisher matrix: general expressions}
\label{sec:FisherGeneral}

The fundamental degrees of freedom in a survey are the positions of the tracers (galaxies, halos or other point-like objects that follow the underlying matter distribution). 
When we measure the number densities $n^i(\vec{x})$ of a tracer species $i$, over some volume around the position $\vec{x}$, that number reflects the mean density of those tracers, $\bar{n}^i(\vec{x})$, as well as the fluctuations $\delta n^i = n^i - \bar{n}^i$.
From these observables we compute the main object that carries information about cosmology, the data (or ``pixel'') covariance:
\begin{equation}
    \label{Eq:DefCorrFun}
    \Gamma^{ij} (\vec{x},\vec{y}) = \langle \delta n^i (\vec{x}) \, \delta n^j (\vec{y}) \rangle 
    = \bar{n}^i (\vec{x}) \,  \bar{n}^j (\vec{y}) \, \xi^{ij}(\vec{x},\vec{y}) + \bar{n}^i (\vec{x}) \delta^{ij} \delta_D (\vec{x} - \vec{y} ) \; ,
\end{equation}
where $\xi^{ij}(\vec{x},\vec{y}) $ is the 2-point correlation function in configuration space, and the last term is shot noise, which we assume here to follow Poisson statistics.
The multi-tracer 2-point correlation function is generally assumed to be related to the matter correlation function, $\xi^{(m)}(\vec{x},\vec{y})$, through some knowable relations such as tracer bias, redshift-space distortions \cite{1987MNRAS.227....1K}, etc. 
In real space (i.e., excluding redshift-space distortion), the matter two-point correlation function can be written in terms of the matter power spectrum as:
\begin{equation}
    \xi^{(m)}(\vec{x},\vec{y}) =
    \xi^{(m)}(|\vec{x} - \vec{y}|) =
    \int \frac{d^3 k}{(2\pi)^3} \, e^{-i \vec{k} \cdot (\vec{x}-\vec{y})} \, P^{(m)}(k) \; .
\end{equation}

Let's say we want to estimate the constraints on some set of parameters $\theta^\mu$, which can be derived from the two-point correlation functions. 
We start by constructing the Fisher information matrix for that set of parameters \cite{tegmark1997karhunen}:
\begin{eqnarray}
    \label{Eq:DefFish}
    F[\theta^\mu , \theta^\nu] = F^{\mu\nu} &=& 
    \frac12 {\rm Tr} \left[ 
    \frac{\partial \Gamma}{\partial \theta^\mu} \Gamma^{-1}
    \frac{\partial \Gamma}{\partial \theta^\nu} \Gamma^{-1}
    \right] \\ \nonumber
    &=& \frac12 \sum_{iji'j'} 
    \int d^3 x_i \int d^3 x_j \int d^3 x_{i'} \int d^3 x_{j'}
    \\ \nonumber
    & & \times \left\{ 
    \frac{\partial \Gamma^{ij}(\vec{x}_i,\vec{x}_j)}{\partial \theta^\mu} [\Gamma^{ji'} (\vec{x}_j,\vec{x}_{i'})]^{-1}
    \frac{\partial \Gamma^{i'j'} (\vec{x}_{i'},\vec{x}_{j'})}{\partial \theta^\nu} [\Gamma^{j'i} (\vec{x}_{j'},\vec{x}_i)]^{-1}
    \right\} \; ,
\end{eqnarray}
where the compact notation for the trace in the definition of the first line denotes both a sum over all tracers, as well as spatial integrations over all positions.
This expression also gives us the first hints that the tracer indices appear linked to the spatial positions where we measure those same tracers: $i \leftrightarrow \vec{x}_i$, $j \leftrightarrow \vec{x}_j$, and so on and so forth. 
This means that we can make use, whenever convenient, of a compact notation whereby the sum over a tracer index also denotes an integral over its corresponding positions in space.

Finally, after computing the Fisher matrix we can invert it to find the covariance matrix for the parameters $\theta^\mu$:
\begin{equation}
    {\rm Cov} [\theta^\mu,\theta^\nu] =
    {\rm Cov}_{\mu\nu} = \left[ F_{\mu\nu} \right]^{-1} \; ,
\end{equation}
which is an estimate of the marginalized uncertainties and degeneracies between the parameters.

\subsection{Fisher matrix in Fourier space}

Working directly in Fourier space allows us to introduce some notions and notation that will become useful later when we compute the multi-tracer Fisher matrix in harmonic space.
For simplicity, in this section we will consider the density contrast for the tracers, $\delta^i = (n^i-\bar{n}^i)/\bar{n}^i$, as our data set.
The Fourier mode $\vec{k}$ of the density contrast  can be expressed as:
\begin{equation}
\label{eqn:dof}
    d^i_a (\vec{k}) = \{ \tilde\delta^i (\vec{k}) \, , \, \tilde\delta^{i*} (\vec{k}) \} \; ,
\end{equation}
where $i=1,2,\ldots,N_t$ denotes the tracer, and $a=1,2$ for the Fourier mode and its complex conjugate, respectively.
The data covariance is then:
\begin{equation}
\label{eqn:expval}
    \langle d^i_a (\vec{k}) d^j_b (\vec{k}{}') \rangle =  D_{ab}  \, \Gamma^{ij} (\vec{k},\vec{k}{}') = \Gamma_{ab}^{ij} (\vec{k},\vec{k}{}')\; ,
\end{equation}
where $D_{ab} = 1-\delta_{ab}$. 
The data covariance in Fourier space is simply the observed power spectrum, including shot noise:
\begin{equation}
\label{eqn:expval2}
\Gamma^{ij} (\vec{k},\vec{k}{}') 
=  \delta_{\vec{k} \, \vec{k}{}'}  \left(  P^{ij} + \frac{\delta_{ij}}{ \bar{n}^i} \right) \; ,
\end{equation}
where $P^{ij} (\vec{k})$ is the power spectrum for the tracers $i,j$.

The Fisher matrix for the set of parameters $\theta^\mu$ is given by the generalized trace:
\begin{equation}
\label{eqn:FishMat}
    F_{\mu\nu} = \frac{1}{4} \sum_k V \tilde{V}_k \sum_{iji'j'} \sum_{aba'b'} 
    \frac{\partial \, \Gamma_{ab}^{ij}}{\partial \theta^\mu}
    \left[ \Gamma_{ba'}^{ji'} \right]^{-1} 
    \frac{\partial \, \Gamma_{a'b'}^{i'j'}}{\partial \theta^\nu} 
    \left[ \Gamma_{b'a}^{j'i} \right]^{-1} \; ,
\end{equation}
where $V$ is the survey volume, $\tilde{V}_k$ is the volume in Fourier space of the bandpowers (Fourier bins) $k$, and the additional factor of $1/2$ in Eq. (\ref{eqn:FishMat}) is due to the fact that our degrees of freedom in Eq. \eqref{eqn:dof} count the Fourier modes twice.

Now, using the fact that the data covariance is separable,
$\left[ \Gamma_{ab}^{ij}\right]^{-1} = \left[ D_{ab}\right]^{-1} \left[ \Gamma^{ij}\right]^{-1}$, and using that $\left[ D_{ab}\right]^{-1} = D_{ab}$, we obtain that $\sum_{abcd} D_{ab} D_{bc} D_{cd} D_{da} = \sum_{ac} \delta_{ac} \delta_{ca} = 2$, hence:
\begin{equation}
\label{eqn:FishMat2}
    F_{\mu\nu} = \frac12 \sum_k 
    V \tilde{V}_k  \sum_{iji'j'}
    \frac{\partial \, \Gamma^{ij}}{\partial \theta^\mu}
    \left[ \Gamma^{ji'} \right]^{-1} 
    \frac{\partial \, \Gamma^{i'j'}}{\partial \theta^\nu} 
    \left[ \Gamma^{j'i} \right]^{-1} \; ,
\end{equation}
In Fourier space the inverse of the data covariance has a trivial expression, namely:
$$
\left[\Gamma^{ij}\right]^{-1} = 
\bar{n}^i \, \delta_{ij} - 
\bar{n}^i \, 
\frac{P^{ij}}{1+{\cal{P}}} 
\, \bar{n}^j \; ,
$$
with ${\cal{P}} = \sum_i \bar{n}^i \, P^{ii}$. 

We now set the parameters $\theta^\mu$ to be the auto- and cross-spectra of the tracers  evaluated at some bandpower, $P^{ij}(k)$.
But before we compute that Fisher matrix, it is worth noting that these spectra (as well as the corresponding data covariance) are symmetric, $P^{ij}=P^{ji}$. Therefore, if we simply identify the parameters of the Fisher matrix as those spectra, we would be counting the cross-spectra twice. 
We handle this double-counting by defining non-degenerate spectra as:
\begin{equation}
    \label{Eq:sympow}
    P^{[ij]} = \left\{ 
    \begin{array}{cc}
    P^{ij} & i \leq j \\
    0   & i > j
    \end{array}
    \right. \; .
\end{equation}
Conversely, we can rewrite the spectra in terms of the non-degenerate spectra as:
\begin{equation}
    \label{Eq:ndpow}
    P^{ij} = P^{[ij]} + P^{[ji]} - \delta_{ij} P^{ii} \; ,
\end{equation}
and a similar expression for the data covariance.
We can now evaluate the partial derivatives assuming that the parameters are the non-denegerate spectra:
\begin{equation}
\label{Eq:InvCovk}
    \frac{\partial \Gamma^{ij}(k)}{\partial P^{[i'j']}(k')} = \delta_{k,k'} \left[ \delta_{ii'}\delta_{jj'}
+ \delta_{ij'}\delta_{ji'} - \delta_{ij}\delta_{ji'}\delta_{i'j'}\delta_{j'i} \right]
\, \equiv \,  \delta_{k,k'} \delta_{[ij],[i'j']}\; .
\end{equation}
Substituting this identity into Eq. (\ref{eqn:FishMat2}) results in a Fisher matrix which is diagonal in the bandpowers, and can be expressed as \cite{Raul2021arXiv211201812A}:
\begin{equation}
\label{eqn:FishMat3}
    F[P^{[ij]},P^{[i'j']}] =
    F^{[ij],[i'j']} = \frac{V \tilde{V}_k }{4} \,
    \left( 2 - \delta_{ij} \right) 
    \left( 2 - \delta_{i'j'} \right) \,
    \left( [\Gamma^{ii'}]^{-1} [\Gamma^{jj'}]^{-1} + [\Gamma^{ij'}]^{-1} [\Gamma^{ji'}]^{-1}  \right)
    \, .
\end{equation}

Eq. \eqref{eqn:FishMat3} makes it clear that, in order to compute the Fisher matrix, we need first to invert the data covariance $\Gamma^{ij}$.
The reason we compute the Fisher matrix for the power spectrum, but not for the correlation function, is that we can, under some approximations, invert the Fourier space data covariance \cite{Hamilton1997,Abramo2010PhRvD}, whereas it is unfeasible to invert the configuration-space data covariance.
As we will show in the next Sections, it is possible to invert the harmonic space data covariance as well, in real and in redshift space. 
Ultimately, this is what has allowed us to compute the Fisher matrix for the angular power spectrum.

Using this notation it is straightforward to show that the inverse of the Fisher matrix of Eq. \eqref{eqn:FishMat3} is indeed the familiar expression for the covariance of the spectra which follows from the 4-point function under the assumption of Gaussianity:
\begin{equation}
    \label{eq:speccov}
    {\rm Cov} [P^{[ij]},P^{[i'j']}] =
    {\rm Cov}^{[ij],[i'j']} =
     \frac{1}{V \tilde{V}_k }\,
    \left( \Gamma^{ii'} \Gamma^{jj'} + \Gamma^{ij'} \Gamma^{ji'}  \right)
    \, .
\end{equation}
In order to show that Eq. \eqref{eq:speccov} is in fact the inverse of Eq. \eqref{eqn:FishMat3}, the following identity is useful:
\begin{eqnarray}
\nonumber
    2 \sum_{[mn]} 
    [\Gamma^{im}]^{-1} \Gamma^{mi'} [\Gamma^{jn}]^{-1} \Gamma^{nj'} 
    &=& 
    \sum_{mn} 
    [\Gamma^{im}]^{-1} \Gamma^{mi'} [\Gamma^{jn}]^{-1} \Gamma^{nj'}
    +
    \sum_{m} 
    [\Gamma^{im}]^{-1} \Gamma^{mi'} [\Gamma^{jm}]^{-1} \Gamma^{mj'}
    \\ \nonumber
    &=& 
    \delta_{ii'} \delta_{jj'}
    +
    \sum_{m} 
    [\Gamma^{im}]^{-1} \Gamma^{mi'} [\Gamma^{jm}]^{-1} \Gamma^{mj'}
    \; .
    \label{eqn:usefulid}
\end{eqnarray}
With the help of this expression it is then straightforward to show that:
\begin{equation}
    \label{Eq:FishCov}
    \sum_{[mn]} F^{[ij],[mn]} \, {\rm Cov}^{[mn],[i'j']} 
    = \delta_{[ij],[i'j']} \, ,
\end{equation}
where we used the same notation as in Eq. \eqref{Eq:InvCovk}.

\subsection{Correlations in harmonic space: the angular power spectrum}

The problem with the Fourier expressions found in the last Section is that, first,
we do not observe the Universe in snapshots ($t=constant$ hypersurfaces), but over the past light-cone. 
And second, the surfaces of constant time in that light-cone are not flat, but 2D spherical shells.
This means that, in order not to miss any physics that depends on evolution, redshift-space effects, or that is only manifested on the largest observable scales, we need a description in terms of the distance over the light cone (or, equivalently, redshift) and spherical harmonics.
This representation can be used both for the density fields (which are scalars under rotations) \cite{Binney1991,HeavensTaylor1995,3DEX} as well as for higher-spin fields such as cosmic shear or flexions \cite{2005PhRvD..72b3516C}.

We wish to express the number of galaxies  of type $i$ that occupy some spherical shell, with radial distances inside the interval $x \in [\bar{x}_i - \Delta x/2 , \bar{x}_i + \Delta x/2 ]$, in terms of an expansion over spherical harmonic functions. 
We then define:
\begin{equation}
    \label{Eq:DefNiEll}
    N^i_{\ell m} (\bar{x}_i) 
    = \int_{\bar{x}_i} d^3 x \, Y_{\ell m}^* (\hat{x}) \, n^i (\vec{x}) 
    = \int_{\bar{x}_i-\Delta x/2}^{\bar{x}_i+\Delta x/2} dx \, x^2 \, \int d^2 \hat{x} \,  Y_{\ell m}^* (\hat{x}) \, n^i (\vec{x}) \; ,
\end{equation}
where $\hat{x}=\vec{x}/x$.
It is useful to define the angular density of tracers in the radial bin $\bar{x}_i$, or {\it counts in shells}, as:
\begin{eqnarray}
    \label{eq:DefNi}
    N^i_{\bar{x}_i} (\hat{x}) 
    &=& \int_{\bar{x}_i} dx \, x^2 \, n^i (x \, \hat{x}) 
    = \int_{\bar{x}_i} dx \, x^2 \, 
    \sum_{p=1}^{N^i_{Tot}} \delta_D (\vec{x}{\,}_i^p - x \, \hat{x}) \\ \nonumber
    &=& \int_{\bar{x}_i} dx \, x^2 \, 
    \sum_{p=1}^{N^i_{Tot}} \, \frac{\delta_D (x^p_i - x)}{x^2} \, \delta_D(\hat{x}^p_i - \hat{x})
    = \sum_{q=1}^{N^i_{\bar{x}_i,Tot}} \delta_D (\hat{x}^q_i -  \hat{x}) \; ,
\end{eqnarray}
where $N^i_{Tot}$ denotes the total number of tracers of type $i$, each one at a position $\vec{x}{\,}_i^p$, and $N^i_{\bar{x}_i,Tot}$ denotes the number of those tracers that fall inside the radial bin $\bar{x}_i$. 
Notice that we have expressed the Dirac delta function for the angular coordinates as $\delta_D(\hat{x}-\hat{x}{}') = \delta_D (\cos\theta - \cos\theta') \, \delta_D (\varphi-\varphi')$, and that the total number of tracers in the radial bin can also be found by integrating the angular density of tracers over the solid angle, 
$N^i_{\bar{x}_i,Tot} = \int d^2 \hat{x} \, N^i_{\bar{x}_i}$.
With these definitions we can rewrite Eq. (\ref{Eq:DefNiEll}) as:
\begin{equation}
    \label{Eq:DefNiEll2}
    N^i_{\ell m} (\bar{x}_i) 
    =  \int d^2 \hat{x} \,  Y_{\ell m}^*  (\hat{x}) \, N^i_{\bar{x}} (\hat{x})
    = \sum_{q=1}^{N^i_{\bar{x}_i,Tot}} Y_{\ell m}^* (\hat{x}^q_i) \; ,
\end{equation}
although we will not make use of the last expression above.
Notice that, with these definitions, and assuming a full sky, the monopole of the number of galaxies in the bin $\bar{x}$ is $N^i_{00} (\bar{x}_i) = N^i_{\bar{x}_i,Tot}/\sqrt{4\pi}$.

Let's now define the harmonic space data covariance as the expectation value of the differential counts of tracers in the shells of radii $\bar{x}$ and $\bar{y}$:
\begin{eqnarray}
    \label{Eq:DefCl}
    \left\langle \delta N^i_{\ell m} (\bar{x}_i) \, \delta N^{j*}_{\ell' m'} (\bar{x}_j) \right\rangle &=&
    \Gamma^{ij}_\ell (\bar{x}_i,\bar{x}_j) 
    \, \delta_{\ell \, \ell'} \, \delta_{m \, m'} 
    \\ \nonumber
    &=& \int_{\bar{x}_i} dx_i \, x_i^2 
    \int_{\bar{x}_j} dx_j \, x_j^2 
    \int d^2 \hat{x}_i \int d^2 \hat{x}_j \,
    Y_{\ell m}^*(\hat{x}_i) Y_{\ell' m'}(\hat{x}_j) \,
    \Gamma^{ij} (\vec{x}_i,\vec{x}_j) 
    \; ,
\end{eqnarray}
where symmetry under rotations leads to the diagonal structure of the angular power spectrum\footnote{The couplings between different $(\ell,m)$ generated by masks or non-trivial angular selection functions \cite{HeavensTaylor1995,Blake2004MNRAS.351..923B} will be the subject of a future paper. 
We also assume in this paper that the radial selection function is constant inside each radial bin, and that the radial bins do not overlap.}.
Notice that the harmonic data covariance above reduces to the angular power spectrum only after subtracting shot noise, $\Gamma_\ell^{ij}(\bar{x}_i,\bar{x}_j) \to C_\ell^{ij}(\bar{x}_i,\bar{x}_j)$.

The expression above relates the data covariance in configuration space, $\Gamma^{ij}(\vec{x}_i,\vec{x}_j)$, to the harmonic data covariance, $\Gamma_\ell^{ij}(\bar{x}_i,\bar{x}_j)$, and it can be inverted in the sense:
\begin{eqnarray}
    \label{Eq:ClCij}
    \Gamma^{ij} (\hat{x}_i \, \bar{x}_i,\hat{x}_j \, \bar{x}_j)  &=& \frac{4\pi}{\Delta V_{\bar{x}_i}} 
    \frac{4\pi}{\Delta V_{\bar{x}_j}}
    \int_{\bar{x}_i} dx_i \, x_i^2 
    \int_{\bar{x}_j} dx_j \, x_j^2 
    \, \Gamma^{ij} (\vec{x}_i,\vec{x}_j)  \\ \nonumber
    &=&
    \frac{4\pi}{\Delta V_{\bar{x}_i}} 
    \frac{4\pi}{\Delta V_{\bar{x}_j}}
    \sum_{\ell m} 
    Y_{\ell m} (\hat{x}_i) \, Y_{\ell m}^* (\hat{x}_j) \,
    \Gamma^{ij}_\ell (\bar{x}_i,\bar{x}_j) 
    \\ \nonumber
    &=&
    \frac{4\pi}{\Delta V_{\bar{x}_i}} 
    \frac{4\pi}{\Delta V_{\bar{x}_j}}
    \sum_{\ell} \frac{2\ell+1}{4\pi}
    {\cal{L}}_\ell(\hat{x}_i \cdot \hat{x}_j) \,
    \Gamma^{ij}_\ell (\bar{x}_i,\bar{x}_j) 
    \; ,
\end{eqnarray}
where the volume of the spherical shell around $\bar{x}$ is given by $\Delta V_{\bar{x}} = \int_{\bar{x}} d^3 x = 4\pi \, \bar{x}^2 \, \Delta x$.
The last line serves to make it explicit the fact that the correlation function in configuration space depends on the angles only through the combination $\mu_{ij} = \hat{x}_i \cdot \hat{x}_j$ that appear in the argument of the Legendre polynomial ${\cal{L}}_\ell$.
Notice also that, in contrast to the configuration-space covariance, which was defined in terms of the number densities $n^i$, the harmonic-space covariance of Eq. \eqref{Eq:DefCl} is defined in terms of the {\em counts in shells}. As a result, whereas the configuration space data covariance, $\Gamma^{ij}(\vec{x}_i,\vec{x}_j)$, has dimensions of 1/(volume)$^2$, the harmonic space data covariance in spherical shells, $\Gamma^{ij}_\ell(\bar{x}_i,\bar{x}_j)$, is adimensional.

Let's say that we are able to find an expression for the inverse of the data covariance in harmonic space, such that:
\begin{equation}
    \label{Eq:InvCell}
    \sum_j \sum_{\bar{y}}
    \left[ \Gamma^{ij}_\ell(\bar{x},\bar{x}_j) \right]^{-1}
    \Gamma^{ji'}_\ell(\bar{x}_j,\bar{x}_{i'}) =
    \delta_{ii'} \delta_{\bar{x}_i \, \bar{x}_{i'}} \; .
\end{equation}
Given such an expression, we could define the inverse of the data covariance in configuration space in analogy to Eq. (\ref{Eq:ClCij}), as:
\begin{equation}
    \label{Eq:InvClCij}
    \left[ \Gamma^{ij} (\hat{x}_i \, \bar{x}_i,\hat{x}_j \, \bar{x}_j) \right]^{-1}
    =
    \frac{\Delta V_{\bar{x}_i}}{4\pi}
    \frac{\Delta V_{\bar{x}_j}}{4\pi}
    \sum_{\ell m} 
    Y_{\ell m} (\hat{x}_i) \, Y_{\ell m}^* (\hat{x}_j)
    \left[ \Gamma^{ij}_\ell (\bar{x}_i,\bar{x}_j) \right]^{-1} \; .
\end{equation}
One can then take the limit back to the continuum by making the volume elements infinitesimally small.
It is then trivial to show, with the help of Eqs. \eqref{Eq:ClCij}-\eqref{Eq:InvCell}, that the expression above is indeed the inverse of the covariance:
\begin{equation}
    \label{Eq:VerifyInv}
    \sum_j \int d^3 x_j 
    \left[ \Gamma^{ij} (\vec{x}_i ,\vec{x}_j) \right]^{-1}
    \Gamma^{ji'} (\vec{x}_j ,\vec{x}_{i'}) = 
    \delta_{ii'} \delta_D (\vec{x}_i - \vec{x}_{i'}) \; ,
\end{equation}
or, equivalently, in finite spherical shells:
\begin{equation}
    \label{Eq:VerifyInvBins}
    \sum_j \sum_{\bar{x}_j} \Delta V_{\bar{x}_j} 
    \left[ \Gamma^{ij} (\hat{x}_i \, \bar{x}_i,\hat{x}_j \, \bar{x}_j) \right]^{-1}
    \Gamma^{ji'} (\hat{x}_j \, \bar{x}_j ,\hat{x}_{i'} \, \bar{x}_{i'}) = 
    \delta_{ii'} \frac{\delta_{\bar{x}_i \, \bar{x}_{i'}}}{\Delta V_{\bar{x}_i}} \,
    \delta_D (\hat{x}_i - \hat{x}_{i'}) \; ,
\end{equation}
where we made use of the identity:
$$
\sum_{\ell m} Y_{\ell m} (\hat{x}) Y_{\ell m}^* (\hat{y}) = \delta_D (\hat{x} - \hat{y}) \; .
$$

We can now go back to the expression for the Fisher matrix given in Eq. (\ref{Eq:DefFish}), and substitute the covariances in configuration space as well as their inverses, Eqs. (\ref{Eq:ClCij}) and (\ref{Eq:InvClCij}). After a bit of algebra we obtain:
\begin{eqnarray}
    \label{Eq:FishClGen}
    F^{\mu\nu} &=& \frac12 \sum_{\ell} (2\ell + 1)
    \sum_{i,\bar{x}_i}
    \sum_{j,\bar{x}_j}
    \sum_{i',\bar{x}_{i'}}
    \sum_{j,\bar{x}_{j'}}
    \frac{\Delta V_{\bar{x}_i}}{4\pi}
    \frac{\Delta V_{\bar{x}_j}}{4\pi}
    \frac{\Delta V_{\bar{x}_{i'}}}{4\pi}
    \frac{\Delta V_{\bar{x}_{j'}}}{4\pi}
    \\ \nonumber
    & & \times
    \frac{\partial \Gamma^{ij}_\ell (\bar{x}_i,\bar{x}_j)}{\partial \theta^\mu}
    \left[ \Gamma^{ji'}_\ell (\bar{x}_j,\bar{x}_{i'}) \right]^{-1}
    \frac{\partial \Gamma^{i'j'}_\ell (\bar{x}_{i'},\bar{x}_{j'})}{\partial \theta^\nu}
    \left[ \Gamma^{j'i}_\ell (\bar{x}_{j'},\bar{x}_i) \right]^{-1} \; .
\end{eqnarray}
In deriving the expression above, the angular integrals cancel all the dependence on the spherical harmonic functions, such that only one sum over $(\ell,m)$ remains, which leads to the factor $\sum_{m=-\ell}^\ell = 2\ell+1$ since none of the terms in the final expression depends on $m$.

What the result above shows is that the Fisher matrix can be decomposed in linearly independent spherical harmonic components:
\begin{equation}
    \label{Eq:FishDecells}
    F^{\mu\nu} = \sum_{\ell} F^{\mu\nu}_\ell \; ,
\end{equation}
where we define:
\begin{equation}
    \label{Eq:FishEllGen}
    F_\ell^{\mu\nu} = \frac{2\ell+1}{2} \, {\rm Tr}
    \left\{
    \frac{\partial \Gamma^{ij}_\ell }{\partial \theta^\mu}
    \left[ \Gamma^{ji'}_\ell \right]^{-1}
    \frac{\partial \Gamma^{i'j'}_\ell }{\partial \theta^\nu}
    \left[ \Gamma^{j'i}_\ell \right]^{-1}
    \right\}
    \; ,
\end{equation}
with the trace denoting, in short-hand notation, both sums over tracer indices as well as integrals over the corresponding radii, $\sum_{\bar{x}_i} \Delta V_{\bar{x}_i}/4\pi (\cdots) \to \int dx_i \, x_i^2 (\cdots)$.
Since we will choose our parameters $\theta^\mu \to C^{ij}_\ell$, and $\partial C^{ij}_\ell / \partial C^{i'j'}_{\ell'} \sim \delta_{\ell \ell'}$, the Fisher matrix for the angular power spectrum is given by Eq. (\ref{Eq:FishEllGen}). 
With partial sky coverage $f_{sky} = \Delta \Omega/4\pi$, where $\Delta \Omega$ is the survey angular area, and collecting different $\ell$'s inside a bin $\bar{\ell}$, the expression becomes \cite{HuJain}:
\begin{equation}
    \label{Eq:FishEllGen2}
    F_{\bar{\ell}}^{\mu\nu} = \frac{f_{sky}}{2} \, \sum_{\ell \in \bar{\ell}} 
    \, (2\ell+1) \, {\rm Tr}
    \left\{
    \frac{\partial \Gamma^{ij}_\ell }{\partial \theta^\mu}
    \left[ \Gamma^{ji'}_\ell \right]^{-1}
    \frac{\partial \Gamma^{i'j'}_\ell }{\partial \theta^\nu}
    \left[ \Gamma^{j'i}_\ell \right]^{-1}
    \right\}
    \; .
\end{equation}

\subsection{Fisher matrix in harmonic space}

Let's now say that we are able to invert the harmonic data covariance (i.e., the observed angular power spectrum), in real or in redshift space. We can then return to Eq. (\ref{Eq:FishEllGen}) and start writing the Fisher matrix for the observables of interest, which in our case are the angular power spectra, $\theta^\mu \to C^{ij}_\ell(\bar{x},\bar{y})$.

As was the case for the power spectrum in Fourier space, we must be aware of the potential for double counting the degrees of freedom. 
For that reason, in analogy with Eq. (\ref{Eq:sympow}) we define:
\begin{equation}
    \label{Eq:sympow2}
    C^{[ij]}_\ell (\bar{x}_i,\bar{x}_j) 
    = \left\{ 
    \begin{array}{cc}
    C^{ij}_\ell (\bar{x}_i,\bar{x}_j) & i \leq j \\
    0   & i > j
    \end{array}
    \right. \; , 
\end{equation}
and
\begin{equation}
    C_\ell^{ij} (\bar{x}_i,\bar{x}_j)    =    C_\ell^{[ij]} (\bar{x}_i,\bar{x}_j) + C_\ell^{[ji]} (\bar{x}_j,\bar{x}_i) - \delta_{ij} C^{ii} (\bar{x}_i,\bar{x}_j) \; ,
\end{equation}
where the auto-correlation in the last term is symmetric in the spatial indices by definition, $C_\ell^{ii}(\bar{x}_i,\bar{x}_j) = C_\ell^{ii}(\bar{x}_j,\bar{x}_i)$.

The partial derivatives in the Fisher matrix in the present case are given by the analog of Eq. (\ref{Eq:InvCovk}):
\begin{equation}
\nonumber
    \frac{\partial C_\ell^{ij} (\bar{x}_i,\bar{x}_j)}{\partial C_\ell^{[i'j']} (\bar{x}_{i'},\bar{x}_{j'})}
    =
    \delta_{ii'} \delta_{jj'} \, \delta_{\bar{x}_i\bar{x}_{i'}}
    \delta_{\bar{x}_j\bar{x}_{j'}}
    + \delta_{ij'} \delta_{ji'} \, \delta_{\bar{x}_i\bar{x}_{j'}}
    \delta_{\bar{x}_j \bar{x}_{i'}}
    - \delta_{ij} \delta_{ji'} \delta_{i'j'}\delta_{j'i} \,  \delta_{\bar{x}_i\bar{x}_j}
    \delta_{\bar{x}_j\bar{x}_{i'}}
    \delta_{\bar{x}_{i'}\bar{x}_{j'}}
    \delta_{\bar{x}_{j'}\bar{x}_i} \; .
\end{equation}
At this point it should have become abundantly clear that the tracer index always appears tied up with the radial bin: $\{i,\bar{x}_i\}$, $\{j,\bar{x}_j\}$, etc. 
Hence, we will often use this association in order to shorten our notation, by implicitly assuming that the two are connected.
In particular, we can define $\bar\delta_{ii'} = \delta_{ii'} \delta_{\bar{x}_i\bar{x}_{i'}}$, leading to: 
\begin{equation}
    \frac{\partial C_\ell^{ij} }{\partial C_\ell^{[i'j']} }
    = 
    \bar\delta_{ii'} \bar\delta_{jj'} 
    + \bar\delta_{ij'} \bar\delta_{ji'}
    - \bar\delta_{ij} \bar\delta_{ji'} \bar\delta_{i'j'} \bar\delta_{j'i} \; .
    \label{eqn:partdev2}
\end{equation}
Substituting this expression into Eq. (\ref{Eq:FishEllGen}) leads us to:
\begin{eqnarray}
\label{Eq:fisherl}
  F[C^{[ij]}_\ell,C^{[i'j']}_\ell] 
  &=& \frac{2\ell+1}{4}
  (2-\bar\delta_{ij}) (2-\bar\delta_{i'j'})
   \left\{
   [{\Gamma}^{i i'}_{\ell}]^{-1}
   [{\Gamma}^{j j'}_{\ell}]^{-1}
   +
   [{\Gamma}^{i j'}_{\ell}]^{-1}
   [{\Gamma}^{i' j}_{\ell}]^{-1}
  \right\}   \; ,
\end{eqnarray}
in full analogy with the result in Fourier space.
And again, in exact correspondence with Fourier space, the covariance matrix in harmonic space is given by:
\begin{eqnarray}
\label{Eq:covl}
  {\rm Cov}[C^{[ij]}_\ell,C^{[i'j']}_\ell] 
  &=& \frac{1}{2\ell+1} \left[
   {\Gamma}^{i i'}_{\ell}
   {\Gamma}^{j j'}_{\ell}
   +
   {\Gamma}^{i j'}_{\ell}
   {\Gamma}^{i' j}_{\ell}
  \right]   \; .
\end{eqnarray}

\section{The harmonic Fisher matrix in real space}

In this Section we will derive exact expressions for the Fisher and covariance matrices of the angular power spectrum for two redshift slices, $C_\ell^{ij}(\bar{x}_i,\bar{x}_j)$, under the assumption of linear biasing for the tracers, but without any redshift-space distortions. 
In the Section 4 we will generalize this result to redshift space, but still under the assumption of linear biasing and using the Kaiser approximation \cite{1987MNRAS.227....1K} -- i.e., assuming that redshift-space distortions are in the linear regime of structure formation.

\subsection{The data covariance in harmonic space}

Let's go back to the covariance in harmonic space, Eq. (\ref{Eq:DefCl}), and write it in the form:
\begin{eqnarray}
    \label{Eq:CovHarmonic}
    \Gamma^{ij}_\ell
    & = & \int_{\bar{x}_   i} dx_i \, x_i^2 
    \int_{\bar{x}_j} dx_j \, x_j^2 
    \int d^2 \hat{x}_i \int d^2 \hat{x}_j \,
    Y_{\ell m}(\hat{x}_i) Y_{\ell m}^*(\hat{x}_j)
    \left\langle \delta n^i (\vec{x}_i) \delta n^j (\vec{x}_j) \right\rangle
    \\ \nonumber
        & = & \int_{\bar{x}_i} dx_i \, x_i^2 
    \int_{\bar{x}_j} dx_j \, x_j^2 
    \int d^2 \hat{x}_i \int d^2 \hat{x}_j \,
    Y_{\ell m}(\hat{x}_i) Y_{\ell m}^*(\hat{x}_j)
    \left[ \bar{n}^i (\vec{x}_i) \, \bar{n}^j (\vec{x}_j) \, 
    \xi^{ij} (\vec{x}_i,\vec{x}_j) \right.
    \\ \nonumber
    & & + \left. \delta_{ij} \, \bar{n}^i(\vec{x}_i) \, 
    \delta_D (\vec{x}_i-\vec{x}_j) \right]
     \; ,
\end{eqnarray}
The first term in this expression is the correlation function of the counts in shells, in harmonic space -- i.e., the angular power spectrum.
If the number densities of tracers in the shells vary as a function of the angular position due to masks or selection functions, then the identity above becomes an angular convolution between the correlation function and those masks and selection functions.
For simplicity, here we will assume that the mean numbers of tracers is nearly constant, so that it can be pulled outside the integral over the radial bins: 
\begin{equation}
    \label{Eq:XiellPk}
    C^{ij}_\ell(\bar{x}_i,\bar{x}_j) 
    = \bar{n}^i(\bar{x}_i) \bar{n}^j(\bar{x}_j) 
    \int_{\bar{x}_i} dx_i \, x_i^2
    \int_{\bar{x}_j} dx_j \, x_j^2
    \int d^2\hat{x}_i \, Y_{\ell m}(\hat{x}_i)\int d^2\hat{x}_j \, Y^*_{\ell m}(\hat{x}_j) \, 
    \xi^{ij}(\vec{x}_i,\vec{x}_j) \, .
\end{equation}
Notice also that under the assumption of a trivial angular selection function, the shot noise term in Eq. \eqref{Eq:CovHarmonic} reduces to $\delta_{ij} \, \bar{N}^i_{\bar{x}_i} = \delta_{ij} \, \bar{n}^i_{\bar{x}_i} \, \Delta V_{\bar{x}_i}$. 

Now comes a simple but crucial step that is often overlooked: if we assume that the correlation function is computed at a $t=\,$constant hypersurface, {\it and} we assume homogeneity, then we can write that, for all $|\vec{x}_i| \in \bar{x}_i$ and all $|\vec{x}_j| \in \bar{x}_j$ we have: 
\begin{eqnarray}
    \label{Eq:XiPk}
    \xi^{ij}(\vec{x}_i,\vec{x}_j)
    &=& 
    \int\frac{d^3k}{(2\pi)^3}
    \, e^{i\vec{k}(\vec{x}_i-\vec{x}_j)}
    P^{ij}(\vec{k}) \, .
\end{eqnarray}
Here the reader should bear in mind that the power spectrum is given in terms of the expression $\langle \tilde{\delta}^i (\vec{k}_i|z_i) \tilde{\delta}^{j*} (\vec{k}_j|z_j) \rangle = (2\pi)^3 \delta_D(\vec{k}_i-\vec{k}_j) P^{ij}(\vec{k}_i)$, and therefore the power spectrum includes factors of the tracer bias as well as growth function. 
In particular, in real space and in the linear regime we would have $P^{ij}(\vec{k}) \to b^i(z_i) \, b^j(z_j) \, D(z_i) \, D(z_j) \, P_m(k|z=0)$.

\begin{figure}
    \centering
    \includegraphics[width = 0.95\textwidth]{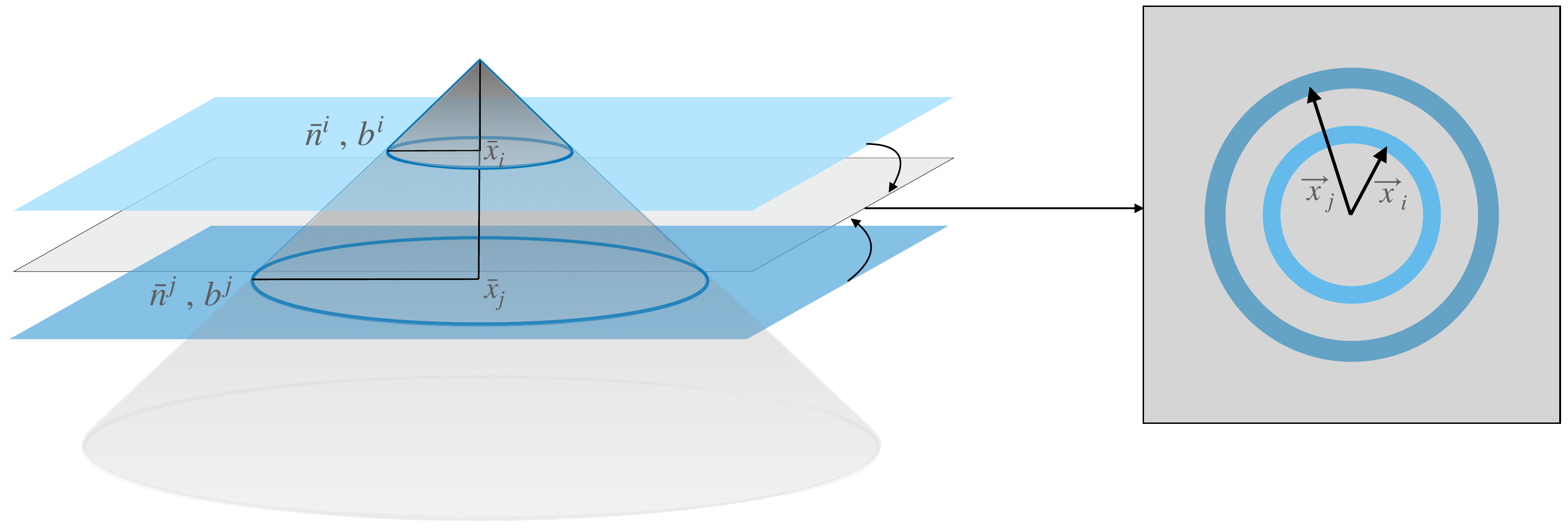}
    \caption{The fields corresponding to the tracer $i$ on the shell $\bar{x}_i$ and to the tracer $j$ on the shell $\bar{x}_j$ lie on the past light-cone of the observer (shown on the left). 
    In order to compute correlation functions (in real or in Fourier space) these fields must be first brought to the same $t=\,$constant hypersurface (shown on the right).
    Homogeneity then allows us to extend the modes inside the annuli $\bar{x}_i$ and $\bar{x}_j$ to the entire hypersurface, and then to compute expectation values such as the correlation function $\xi^{ij}$ and the power spectrum $P^{ij}$. 
    Notice that the quantities $\bar{n}^i$, $b^i$, $\bar{n}^j$ and $b^j$, etc., are kept fixed in the common $t=\,$constant hypersurface.}
    \label{fig:LCs}
\end{figure}

At this point it is important to remember that the wavenumber $\vec{k}$ in the integral above has support in the {\it entire} 3D volume, and not only in the spherical shells $\bar{x}_i$ and $\bar{x}_j$. 
Now, recall that the two spherical shells $z_i$ (of comoving radius $\bar{x}_i$) and $z_j$ (of radius $\bar{x}_j$) are {\it not} in the same $t=\,$constant hypersurfaces -- in fact, the corresponding observables are on the past light-cone of the observer.
Eq. \eqref{Eq:XiPk} is, therefore, the result of  bringing both spherical shells to the same configuration space volume (and the same $t=\,$constant hypersurface), and then taking the expectation value of the 2-point function of the Fourier modes $\langle \tilde{\delta}^i(\vec{k}_i) \tilde{\delta}^{j*}(\vec{k}_j) \rangle$ in that hypersurface -- which is arbitrary, although it is often assumed to be the one at $z=0$.
The procedure is pictured in figure \ref{fig:LCs}.
If we assume that the number densities and biases corresponding to the annuli $\bar{x}_i$ and $\bar{x}_j$ are kept fixed, and we restrict our attention to the linear regime, then it makes no difference what is the common $t=\,$constant hypersurface that we chose to compute the expectation values. 
If, on the other hand, we allow for non-linearities that change as a function of time due to the growth of structures, then in principle the $t=\,$constant hypersurface should be defined carefully -- or, in  a more general approach, one could consider using unequal-time correlation functions \cite{2017PhRvD..95f3522K}.

Coming back to Eq. \eqref{Eq:XiellPk}, substituting Eq. \eqref{Eq:XiPk} and using the Rayleigh expansion of a plane wave in  spherical harmonics:
\begin{equation}
    e^{i\vec{k} \cdot \vec{x}} = 4\pi \sum_{\ell m} i^\ell j_\ell(kx)Y_{\ell m}(\hat{k}) Y_{\ell m}^*(\hat{x}) \; ,
\end{equation}
after a bit of trivial algebra we obtain:
\begin{eqnarray}
    \label{Eq:xi}
    C^{ij}_\ell
    &=&  
    \bar{n}^i \, \bar{n}^j
    \int_{\bar{x}_i} dx_i \, x_i^2
    \int_{\bar{x}_j} dx_j \, x_j^2
    \times
    \frac{2}{\pi}\int_0^\infty dk \, k^2 
    \, j_\ell(kx_i)\, j_\ell(kx_j) \, P^{ij}(k) 
    \, .
\end{eqnarray}
In the linear regime we can simplify this expression further, using $P^{ij}(k) = b^i b^j D^i D^j P^{(m)}(k)$, where $P^{(m)}(k)$ is the matter power spectrum and $D^i = D(z_i)$ is the growth function.
Taking the biases of the tracers outside the volume integral (again a safe assumption, if the slices are thin enough) we obtain:
\begin{eqnarray}
    \nonumber
    C^{ij}_\ell
    &=&  
    \bar{n}^i \bar{n}^j
    \, b^i \, b^j \, D^i \, D^j 
    \int_{\bar{x}_i} dx_i \, x_i^2
    \int_{\bar{x}_j} dx_j \, x_j^2
    \times
    \frac{2}{\pi}\int_0^\infty dk \, k^2 
    \, j_\ell(kx_i)\, j_\ell(kx_j) \, P^{(m)}(k) 
    \\ 
    \label{Eq:xijm}
    &=&  
    \bar{n}^i \bar{n}^j
    \, b^i \, b^j \,
    C^{(m)}_\ell(\bar{x}_i,\bar{x}_j)  \, ,
\end{eqnarray}
where $C^{(m)}_\ell(\bar{x}_i,\bar{x}_j) $ is the matter angular power spectrum for the two slices.
It is also possible to extend this expression to take into account non-linearities -- see, e.g., \cite{2020PhRvD.102h3521G}

Another useful approximation arises if the Bessel functions are nearly constant inside each radial bin: \begin{eqnarray}
    \label{Eq:xijell}
    C^{ij}_\ell(\bar{x}_i,\bar{x}_j) 
    &\approx&  
    \bar{N}^i
    \bar{N}^j
    \times
    \frac{2}{\pi}\int_0^\infty dk \, k^2 
    \, j_\ell(k\bar{x}_i)\, j_\ell(k\bar{x}_j) \, P^{ij}(k) \equiv \bar{C}{}^{ij}_\ell(\bar{x}_i,\bar{x}_j)
    \, ,
\end{eqnarray}
where we defined $\bar{N}^i = \bar{n}^i(\bar{x}_i) \, \Delta V_{\bar{x}_i}/4\pi$ as the mean number of tracers {\em per unit solid angle}, and we assumed that the spherical shells are sufficiently thin, such that we can approximate the argument of the spherical Bessel functions to the mean radius, $x \to \bar{x}$. 
We will {\em not} rely on the approximation of Eq. (\ref{Eq:xijell}) for now -- although it will turn out to be useful later. 
For more details on this approximation we refer the reader to Appendix A.

At this point it is relevant to recall that we are working under the assumption of linear regime.
At $z=0$, scales smaller than about $ \Delta r \lesssim 30 \, h^{-1}$ Mpc (i.e., $k \gtrsim 0.2 \, h$ Mpc$^{-1}$) are already affected by non-linear effects, but at higher redshifts these non-linear scales move in proportion to the matter growth rate.
Therefore, one can think of the minimum radial bin widths where our results would still remain applicable as $\Delta r \gtrsim 30 \, h^{-1}$ Mpc$/(1+z)$, or $\Delta z \gtrsim 0.01/(1+z)$.

\subsection{Inverting the data covariance matrix}

We now return to the task of inverting the data covariance of 
Eq. \eqref{Eq:CovHarmonic}, in light of the expression for the correlation function found in Eq. \eqref{Eq:xi}.
Using the closure relation for spherical Bessel functions,
\begin{equation}
\label{Eq:closure}
    \int_0^\infty dt \, t^2 \, j_\ell(a t) \, j_\ell(b t) = \frac{\pi}{2} \frac{\delta_D(a-b)}{a^2} \; ,
\end{equation}
allows us to rewrite the harmonic data covariance as:
\begin{eqnarray}
    \label{Eq:CovHarmonic2}
    \Gamma^{ij}_\ell
    & = & 
    \bar{n}^i \bar{n}^j
    \int_{\bar{x}_i} dx_i \, x_i^2
    \int_{\bar{x}_j} dx_j \, x_j^2
    \times
    \frac{2}{\pi}\int_0^\infty dk \, k^2 
    \, j_\ell(kx_i)\, j_\ell(kx_j) \, M^{ij}(k) \; ,
\end{eqnarray}
where we defined the matrix:
\begin{equation}
    M^{ij} (k) =  P^{ij}(k)
    + \frac{\delta_{ij}}{ \sqrt{\bar{n}^i \, \bar{n}^j} } \; .
\end{equation}

This last expression indicates how we can go about inverting the harmonic covariance. But before we present that inverse covariance, consider that, in the continuum limit, we have:
$$
\sum_{\bar{x}} \int_{\bar{x}} dx \, f(x) \, g(\bar{x}) 
\to 
\int dx \, f(x) \, g(x) \; .
$$
With the help of this limit, as well as Eq. (\ref{Eq:CovHarmonic2}), it is easy to show that the inverse data covariance is given by:
\begin{equation}
    \label{Eq:InvCovCl}
    \left[ \Gamma^{ji'}_\ell  \right]^{-1} =
     \frac{1}{\bar{n}^j \, \bar{n}^{i'}}
    \times
    \frac{2}{\pi}\int_0^\infty dk \, k^2 
    \, j_\ell(k\bar{x}_j)\, j_\ell(k\bar{x}_{i'}) \, \left[ M^{ji'}(k) \right]^{-1}\; ,
\end{equation}
which then leads to Eq. (\ref{Eq:InvCell}), i.e.:
\begin{equation}
    \label{Eq:CovInvCov}
    \sum_j \sum_{\bar{x}_j} 
    \, \Gamma^{ij}_\ell (\bar{x}_i,\bar{x}_j)\, 
    \left[ \Gamma^{ji'}_\ell (\bar{x}_j,\bar{x}_{i'}) \right]^{-1} =
    \delta_{ii'} \delta_{\bar{x}_i \, \bar{x}_{i'}} \; .
\end{equation}
It is important to recall that, according to the discussion in the previous section, the integration variable above, $\bar{x}_j$, corresponds to the radius in a  $t=\,$constant hypersurface, and the index $j$ corresponds to all tracers in that hypersurface, which means that the tracer number densities $\bar{n}^j$ and biases $b^j$ are kept constant -- i.e., they do not depend on $\bar{x}_j$.

In the definition of the inverse covariance, Eq. \eqref{Eq:InvCovCl}, the inverse of the matrix $M^{ij}$ appears. 
We first rewrite that matrix as:
\begin{eqnarray}
\label{Eq:MatDef}
M^{ij}(k) &=& 
    \frac{1}{\sqrt{\Bar{n}^i }}
    \left[ \delta_{ij} 
    + {\cal{P}}^{ij}(k)
    \right] 
    \frac{1}{\sqrt{\Bar{n}^j}}
    \; ,
\end{eqnarray}
where in the last line we defined ${\cal{P}}^{ij} = \sqrt{\bar{n}^i \, \bar{n}^j} \, P^{ij} $.
Notice also that this adimensional quantity is separable, in the sense that ${\cal{P}}^{ij} {\cal{P}}^{i'j'} = {\cal{P}}^{ii'} {\cal{P}}^{jj'}$, and $\sum_l {\cal{P}}^{il} {\cal{P}}^{jl}= {\cal{P}}^{ij} {\cal{P}} $, where we defined the trace
${\cal{P}} = \sum_i {\cal{P}}^{ii} $.
Due to these properties, the inverse of the matrix $M$ is given simply by:
\begin{eqnarray}
    \label{Eq:InvMatReal}
    [M^{ij}]^{-1} 
    &=&
    \sqrt{\Bar{n}^i}
    \left[ \delta_{ij} 
    - \frac{{\cal{P}}^{ij} }{1+{\cal{P}}}
    \right]
    \sqrt{\Bar{n}^j} \; .
\end{eqnarray}

At this point it is useful to understand the fundamental reason why we were able to invert the harmonic data covariance in real space, and which will later allow us to also invert that covariance including redshift-space distortions. 
First, we rewrite the covariance as:
\begin{eqnarray}
    \label{Eq:Covxxreal}
    \Gamma^{ij}_\ell (\bar{x}_i,\bar{x}_j) 
    & = & 
    \bar{N}^i
    \bar{N}^j
    \; \frac{2}{\pi} \int_0^\infty dk_i \, k_i^2
    \; \frac{2}{\pi} \int_0^\infty dk_j \, k_j^2
    \;
    j_\ell(k_i \bar{x}_i) \, j_\ell(k_j \bar{x}_j)
    \,  \Gamma^{ij}_\ell(k_i,k_j) \,  ,
\end{eqnarray}
where we defined the Fourier conjugate of the data covariance:
\begin{eqnarray}
    \label{Eq:Covkkpreal}
    \Gamma^{ij}_\ell (k_i,k_j) 
    & = & 
    \frac{\pi}{2} \frac{\delta{(k_i-k_j)}}{k_i^2} \left[ 
    \frac{\delta_{ij}}{\bar{n}^i} + P^{ij}(k_i)  \right] \; .
\end{eqnarray}
Since this object is perfectly diagonal in the Fourier modes, it is straightforward to find its inverse -- in stark contrast with the angular power spectrum as a function of the radii of the spherical shells, which cannot be directly inverted. 
Therefore, in Fourier space the problem reduces to finding the inverse of a matrix in the tracer indices, $M^{ij}$, which is completely trivial.
This argument clarifies what we have done above: we first found the inverse of Eq. \eqref{Eq:Covkkpreal}, and then transformed it back to real space, arriving at Eq. \eqref{Eq:InvCovCl}.
We will use a very similar approach later, in Section 4, when we tackle the angular power spectrum in redshift space.

\subsection{An exact solution}

Consider a single tracer with number density $\bar{n}$, unit bias ($b=1$), and matter density fluctuations with a  ``top-hat'' power spectrum $P(k) = P_0 \left[ \theta_H(k-k_1) - \theta_H(k-k_2) \right]$,
where $\theta_H$ is the Heaviside (step) function --- i.e., the spectrum is only non-zero for $k_1 \leq k \leq k_2$. 
To be clear, in this example we neglect any time evolution effects as well as redshift-space distortions.
The Fourier-space data covariance is then by:
\begin{equation}
    \Gamma(k) = V_0 [1 + A \, \theta_H(k-k_1) - A \, \theta_H(k-k_2)] \; ,
\end{equation}
where $V_0 = 1/\bar{n}$ corresponds to the shot noise term, and $A$ is an adimensional parameter, so that the amplitude of the spectrum is $P_0 =A V_0$.

As shown in Appendix B, the harmonic data covariance corresponding to this power spectrum is given by the analytical expressions:
\begin{eqnarray}
\label{eq:1exactcl3}
\Gamma_\ell(\bar{x},\bar{y}) 
&=&
\bar{N}_{\bar{x}} \bar{N}_{\bar{y}} V_0 \left\{ \delta_{\bar{x} \, \bar{y}} 
\frac{4\pi}{\Delta V_{\bar{x}}} + 
    \, \frac{2}{\pi} A \left[ \, 
    k_1^3 \, g_\ell(k_1 \bar{x}, k_1 \bar{y}) 
    -
    k_2^3 \, g_\ell(k_2 \bar{x}, k_2 \bar{y}) \right]
    \right\} \; ,
\\
\label{eq:1exactclinv3}
\Gamma_\ell^{-1} (\bar{x},\bar{y}) 
&=&
V_0 \left\{ \delta_{\bar{x} \, \bar{y}} 
\frac{4\pi}{\Delta V_{\bar{x}}}
    - \, \frac{2}{\pi}\frac{A}{1+A} \, 
    \left[ k_1^3 \, g_\ell(k_1 \bar{x}, k_1 \bar{y}) 
    -  
    k_2^3 \, g_\ell(k_2 \bar{x}, k_2 \bar{y}) \right]
    \right\} 
 \; ,
\end{eqnarray}
where $\bar{N}_{\bar{x}} = \bar{n} \, \Delta V_{\bar{x}}/4\pi$ is the number of tracers per unit solid angle in the radial bin ${\bar{x}}$,
and:
\begin{eqnarray}
    g_\ell (x,y) &\equiv&
    - \frac{x\, j_{\ell-1}(x) j_\ell(y) - y\, j_{\ell-1}(y) j_\ell(x)}{x^2 - y^2} 
    \\ \nonumber
    &=& 
    \frac{ j_0(x-y) - (-1)^\ell \, j_0 (x+y) }{2\, x\,y} 
    - \frac{1}{x\,y} \sum_{n=0}^{n_\ell} [2(\ell-1-2n)+1] \, j_{\ell-2n} (x) \, j_{\ell-2n}(y) \; .
\end{eqnarray}
The sum in the second line above is over 
$n=0,1, \ldots, n_\ell = (2\ell-1)/4$, so that 
$ 2(\ell-1-2n)+1 \geq 0$, so that the sum stops at $n_\ell = (\ell-1)/2$ for odd values of $\ell$, and at $n_\ell = \ell/2$ for even values of $\ell$.
The expression of the second line shows explicitly that this function is well-behaved as $x\to y$, and it is symmetric under $x \leftrightarrow y$, since $j_0(x-y) = j_0(y-x)$.

\begin{figure}
    \centering
    \includegraphics[width = 0.98\textwidth]{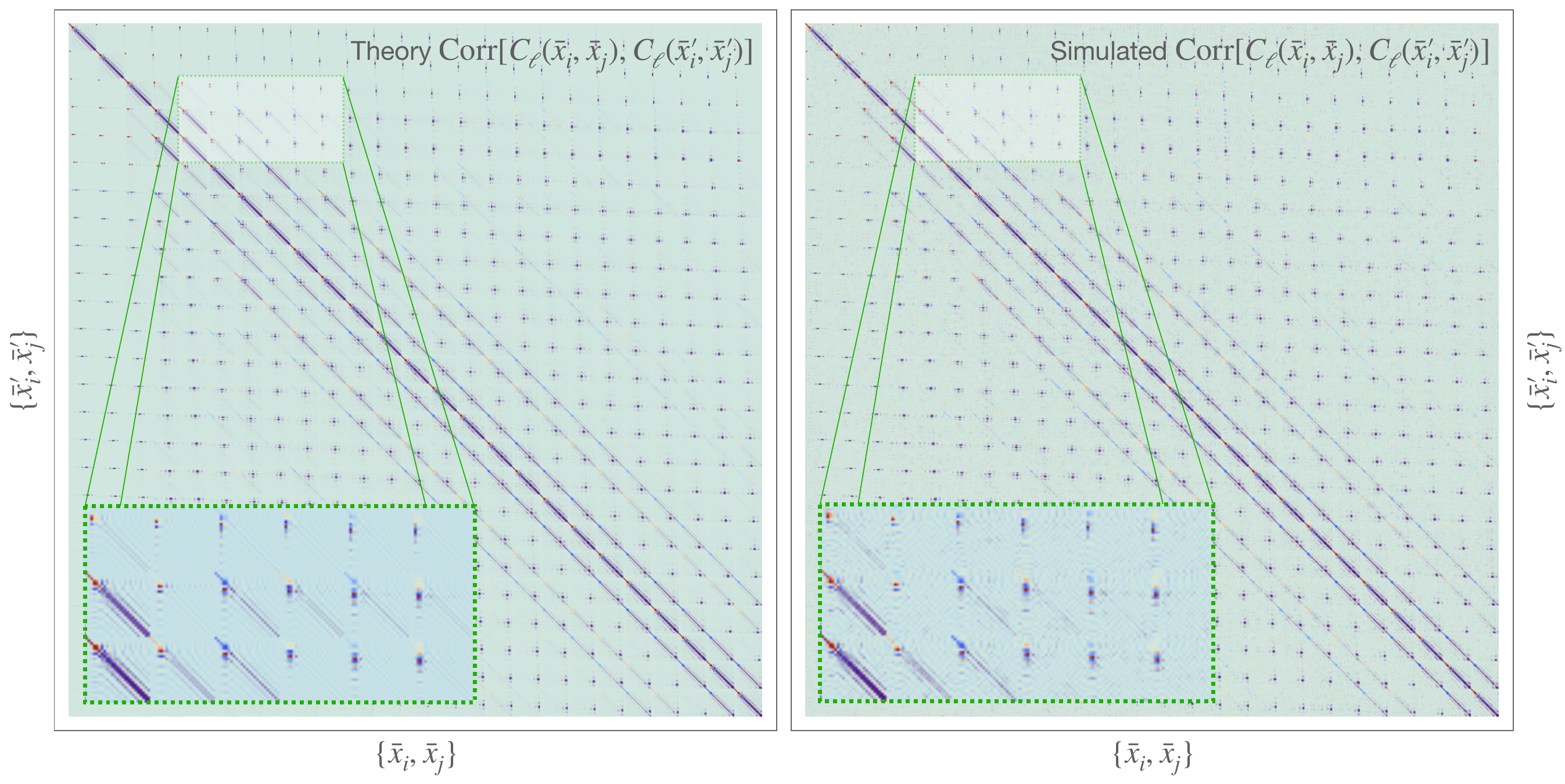}
    \caption{Correlation matrix of the angular power spectrum corresponding to a top-hat Fourier power spectrum, for the multipole $\ell=10$.
    We split the radial coordinate in 25 bins of width 
    $\Delta x = 10\,  h^{-1} \, {\rm Mpc}$ from $\bar{x}_1=5\,  h^{-1} \, {\rm Mpc}$ to $\bar{x}_{25}=245\, h^{-1} \, {\rm Mpc}$, resulting in a $625 \times 625$ covariance matrix. 
    The left panel shows the analytical (``Theory'') correlation matrix computed with the help of  Eq. \eqref{Eq:covl}. 
    The right panel shows the sample correlation from 1000 Gaussian simulations. 
    The insets show a zoom in the same parts of the two matrices, for better comparison between theory and simulations.}
    \label{fig:norm_complete_cov_matrices}
\end{figure}

We can use the expressions in Eqs. \eqref{eq:1exactcl3}-\eqref{eq:1exactclinv3} to compare our
analytical expression for the covariance with Guassian simulations. 
We produced a sample of 1000 simulation boxes using the ``top-hat'' spectrum, and computed the sample mean and sample variance of the angular power spectra.
We use 25 redshift slices corresponding to radial bins of width $\Delta x= 10 \,h^{-1} \, {\rm Mpc}$ from $\bar{x}=5 \,h^{-1} \, {\rm Mpc}$ to $\bar{x}=245\,h^{-1} \, {\rm Mpc}$, and for brevity we focus here on the multipole $\ell=10$.
For more details, see Appendix B.

In figure \ref{fig:norm_complete_cov_matrices} we compare the analytical covariance of Eq. \eqref{Eq:covl} (left panel) with the sample covariance (right panel).
For visualization purposes we show the entire $[N_r\times N_r]\times[N_r\times N_r]$ matrix, where $N_r$ is the number of radial bins from our mock simulations -- i.e., we compute the whole ${\rm Cov}[C_\ell(\bar{x}_i,\bar{x}_j),C_\ell (\bar{x}_{i'},\bar{x}_{j'})]$, even if that carries redundancies due to double-counting the cross-spectra. 
Moreover, in order to further facilitate visualization, we display the correlation matrix, ${\rm Corr}[X,X'] = {\rm Cov}[X,X']/\sqrt{{\rm Cov}[X,X] {\rm Cov}[X',X']}$.
Since we have 25 radial bins, these are $(25^2) \times  (25^2) = 625 \times 625$ correlation matrices.

It can be seen from figure \ref{fig:norm_complete_cov_matrices} that all the main structures of the theoretical correlation matrix (left panel) are reflected in the sample correlation matrix constructed from our 1000 Gaussian simulations (right panel). 
Notice, in particular, the significant levels of correlation (and, therefore, covariance) between different different pairs of slices.
We conclude by pointing out that the exercise above has served to validate the expressions found in previous sections for the angular power spectrum and for the covariance matrix of the power spectra. 
As a consequence, we also indirectly checked our expression for the Fisher matrix of the angular power spectra, Eq. \eqref{Eq:fisherl}.

\section{The harmonic Fisher matrix in redshift space}
    
We will now take into account redshift space distortions \cite{1987MNRAS.227....1K} in the linear regime of structure formation, and subsequently compute the Fisher matrix directly in redshift space.

Analogously to the real space case, the two-point correlation function is defined here as the expectation value of the counts inside the spherical shells of radii $\bar{x}$ and $\bar{y}$, but now including the effects of the peculiar velocities -- see, e.g., \cite{FSL1994MNRAS,1996MNRAS.278...73H,2020PhRvD.102h3521G}:
\begin{eqnarray}
    \label{Eq:rsdxi}
    C_{\ell,s}^{ij}(\bar{x}_i,\bar{x}_j) &=&
    \;    \bar{n}^i_{\bar{x}_i} \, \bar{n}^j_{\bar{x}_j}
    \int_{\bar{x}_i} dx_i \, x_i^2
    \int_{\bar{x}_j} dx_j \, x_j^2
    \\ \nonumber
    & & \quad \times \frac{2}{\pi}\int_0^\infty dk\, k^2
    \left[ j_\ell (kx_i) - \beta^i  j_\ell''(kx_i) \right] \left[ j_\ell (kx_j) - \beta^j j_\ell''(kx_j) \right] P^{ij}(k) \; ,
\end{eqnarray}
with $\beta^i = f(z_i)/b^i$, where $f(z)$ is the matter growth rate. When $\beta^i \to 0$ the effects of the peculiar velocities are erased, and this expression reduces to the real space correlation function given by Eq. \eqref{Eq:xi}. The harmonic data covariance in redshift space is therefore a simple generalization of Eq. \eqref{Eq:CovHarmonic2}: 
\begin{eqnarray}
    \label{Eq:zspace_cov}
    \Gamma_{\ell,s}^{ij}(\bar{x}_i,\bar{x}_j) &=&  
    \bar{n}^i \bar{n}^j
    \int_{\bar{x}_i} dx_i \, x_i^2 \, 
    \int_{\bar{x}_j} dx_j \, x_j^2 \, 
    \frac{2}{\pi}
    \int_0^\infty dk\, k^2 \; \Big\{ 
    j_\ell (kx_i)j_\ell (kx_j) \frac{\delta_{ij}}{ \sqrt{\bar{n}^i \, \bar{n}^j} }
    \\
    \nonumber
    && \; + 
    \left[ j_\ell (kx_i) - \beta^i j_\ell''(kx_i) \right] 
    \left[ j_\ell (kx_j) - \beta^j j_\ell''(kx_j) \right] P^{ij}(k)
    \Big\} \; .
\end{eqnarray}
Just as was the case in real space, the challenge is to invert this data covariance. 
We will show in the next section how to invert Eq. \eqref{Eq:zspace_cov} in the same sense of Eq. \eqref{Eq:InvCell}, that is:
\begin{equation}
    \label{Eq:InvCellredshift}
    \sum_j \sum_{\bar{x}_j}
    \left[ \Gamma^{ij}_{\ell,s}(\bar{x}_i,\bar{x}_j) \right]^{-1}
    \Gamma^{ji'}_{\ell,s}(\bar{x}_j,\bar{x}_{i'}) =
    \delta_{ii'} \delta_{\bar{x}_i \, \bar{x}_{i'}} \; .
\end{equation}

In order to accomplish this inversion, and in analogy to what was done in real space, we will make use of analytical solutions to the integrals of products of spherical Bessel functions that appear in Eq. \eqref{Eq:zspace_cov}.
The term without any derivatives is simply the closure relation, Eq. \eqref{Eq:closure}, already used in the real space case. The challenge lies with the terms involving the second derivatives.

We start by recalling that the spherical Bessel functions satisfy the  differential equation:
\begin{equation}
    \label{4}
     z^2j_\ell''(z)+2zj_\ell'(z)+\big[z^2-\ell(\ell+1)\big]j_\ell(z)=0 \; .
\end{equation}
We can combine this with the recurrence relation:
\begin{equation}
    \label{5}
    j_\ell'(z)=\frac{\ell}{z} j_\ell(z)-j_{\ell+1}(z) \; ,
\end{equation}
and rewrite the second derivative of the spherical Bessel function as:
\begin{equation}
    \label{ddj}
     j_\ell''(z)=\frac{1}{z^2}\bigg\{\big[\ell^2-\ell-z^2\big] j_\ell(z)+2\,z\, j_{\ell+1}(z)\bigg\} \; .
\end{equation}

At this point it is important to draw a distinction between our path and that chosen since Fisher, Scharf and Lahav \cite{FSL1994MNRAS} first wrote an expression for the angular power spectrum in redshift space.
After that seminal paper, many subsequent others, e.g. Ref. \cite{Padmanabhan2007MNRAS.378..852P}, expressed the second derivative of the spherical Bessel function in terms of Bessel functions of different orders:
\begin{equation}
    j_\ell''(z) = \frac{\ell(\ell-1)}{(2\ell+1)(2\ell-1)} j_{\ell-2} (z) + \frac{(\ell+1)(\ell+2)}{(2\ell+1)(2\ell+3)} j_{\ell+2} (z) -
    \frac{(2\ell^2 +2\ell-1)}{(2\ell+3)(2\ell-1)} j_{\ell} (z) \; .
\end{equation}
We, on the other hand, employ the expression given by Eq. \eqref{ddj}, which, as shown in Appendix C, leads to analytical integrals.

There are two relevant integrals, the first one involving the combination of spherical Bessel functions: 
\begin{equation}
\label{H}
    \int dk\, k^2 j_\ell (k{x}) j_\ell''(k{y})  = -\frac{\pi}{2{x}^2}\delta_D({x}-{y}) + H_\ell({x},{y}) \; ,
\end{equation}
where, as shown in Appendix C:
\begin{equation}
\label{eq:Hl}
    H_\ell({x},{y}) \, =  \,
    \frac{\pi}{2y^2}
\begin{dcases}
    \frac{1}{x}\left[\frac{2(2\ell+1)+\ell(\ell-1)}{2(2\ell+1)}\right]& \text{if } {x}={y}\\[12pt]
    \frac{1}{{x}}\left(\frac{{y}}{{x}}\right)^\ell\left[\frac{2(2\ell+1)+\ell(\ell-1)}{(2\ell+1)}\right]& \text{if } {x}>{y} \\[12pt]
    \frac{1}{{y}}\frac{\ell(\ell-1)}{(2\ell+1)}\left(\frac{{x}}{{y}}\right)^\ell & \text{if } {x}<{y} \; .
\end{dcases} 
\end{equation}
Although this expression shows a discontinuity, that does not turn out to be a problem since, as we shall see later on, it appears twice, with interchanged arguments. 
In particular, the symmetric combination of this expression appears when we consider auto-correlations of the same tracer, and in that case the definition above simplifies to a manifestly continuous expression:
\begin{equation}\label{G_rr}
    \frac12 \left[ H_\ell({x},{y}) + H_\ell({y},{x}) \right] = \frac{\pi}{2 \, r_>}\left(\frac{r_<}{r_>}\right)^\ell
    \left[\frac{2(2\ell+1)r_<^2+\ell(\ell-1)({x}^2+{y}^2)}{2(2\ell+1) {x}^2 {y}^2}\right] \; .
\end{equation}

The second relevant integral involves two second derivatives of spherical Bessel functions of the same order, and as also shown in Appendix C we have:
\begin{equation}\label{G}
    \int_0^\infty dk \, k^2 \, j''_\ell(k {x})j''_\ell(k {y})
    = \frac{\pi}{2} \frac{\delta_D( {x}- {y})}{ {x}^2} + G_{\ell}( {x}, {y}) \; ,
\end{equation}
where:
\begin{eqnarray}
    \label{Eq:Intjj}
    G_{\ell}( {x}, {y})
    & \equiv & 
    \frac{\pi}{2} \frac{r_<^\ell}{r_>^{\ell+1}} 
    \left\{ 
    \left[ \frac{4}{(2\ell+3)} -2 \right] \frac{1}{r_>^2} \right.
    \\     \nonumber
    & & + \frac{2 \, \ell(\ell-1)}{(2\ell+1)(2\ell+3)} 
    \frac{1}{r_>^2}    - 
    \frac{\ell(\ell-1)}{(2\ell+1)}
    \left( \frac{1}{r_>^2} + \frac{1}{r_<^2} \right)
    \\
    \nonumber
    & & + \frac{2 \, \ell(\ell-1)}{r^2_< }  \left[ \frac{1}{(2\ell+1)} - \frac{1}{(2\ell+3)} \left( \frac{r_<}{r_>} \right)^2  \right] \\ 
    \nonumber
    & & \left.
    + 2^\ell [\ell(\ell-1)]^2 
     \frac{(\ell+1)!(2\ell-3)!!}{(2\ell+3)!}
     \frac{1}{r_< ^2}
     \left[ 2\ell+3-(2\ell-1)\left( \frac{r_<}{r_>} \right)^2 \right]  \right\} \; .
\end{eqnarray}
The relevance of these analytic integrals will become clear in the next Section.


\subsection{Inverting the data covariance in redshift space}

We are now in a position to show how to invert the expression for the harmonic data covariance in redshift space.
The reader who is uninterested in the details of this calculation can jump to the final results, which are presented in Section 4.2.

We start this calculation by rewriting the data covariance:
\begin{eqnarray}
    \label{Eq:covredspace}
    \Gamma_{\ell,s}^{ij}( \bar{x}_i, \bar{x}_j) &=&  
     \bar{n}^i \, \bar{n}^j
    \int_{\bar{x}_i} dx_i \, x_i^2
    \int_{\bar{x}_j} dx_j \, x_j^2
    \frac{2}{\pi} \int_0^\infty dk\, k^2 \Bigg[ 
    j_\ell (kx_i)j_\ell (kx_j) 
    \frac{\delta_{ij}}{ \sqrt{\bar{n}^i \, \bar{n}^j} }
    \\
    \nonumber
    && \, + \left[ j_\ell (k x_i) - \beta^i j_\ell''(k x_i) \right]  \left[ j_\ell (kx_j) - \beta^j j_\ell''(kx_j) \right] P^{ij}(k)
     \Bigg] 
    \\ \nonumber
    & \approx & 
    \bar{N}^i \, \delta_{ij} \,  \delta_{\bar{x}_i \, \bar{x}_j} 
    \\ \nonumber
    & & \, + \,
    \bar{N}^i \bar{N}^j
    \frac{2}{\pi}
    \int_0^\infty dk\, k^2 \left[ j_\ell (\bar{x}_i) - \beta^i j_\ell''(k \bar{x}_i) \right] 
    \left[ j_\ell (k\bar{x}_j) - \beta^j j_\ell''(k\bar{x}_j) \right] P^{ij}(k) \; ,
\end{eqnarray}
where $\bar{N}^i=\bar{n}^i(\bar{x}_i) \Delta V_{\bar{x}_i}/4\pi$ denotes the mean angular number density of tracers $i$ in the radial bin $\bar{x}_i$. 
Also recall that $\delta_{\bar{x}_i \, \bar{x}_j}=1$ if the two shells (radial bins) are the same, $\bar{x}_i=\bar{x}_j$, and 0 otherwise.
The approximation taken from the first to the second line is similar to that which was discussed in Section 2.1 and was shown in figure \ref{fig:residues}. 
Our results do not rely on this approximation, but we use it here just in order to shorten our expressions.

We will now simplify the nature of our problem by using the condensed notation:
\begin{equation}
    \label{Eq:SomeDefs}
    \frac{\pi}{2}
    \frac{\delta_D(k-k')}{k^2} \rightarrow \delta^{kk'}_F    \qquad , \qquad 
    \frac{2}{\pi} \int dk \, k^2 \rightarrow \sum_{[k]} \; ,
\end{equation}
and define the normalized data covariance:
\begin{equation}
    \label{Eq:NormFac}
    \Gamma^{ij}_{\ell,s} \quad \rightarrow \quad
    \tilde{\Gamma}^{ij}_{\ell,s} \, \equiv \,
    \frac{1}{\sqrt{\bar{N}^i \, \bar{N}^j}} \, \Gamma^{ij}_{\ell,s} \; .
\end{equation}

Now let's define the auxiliary ``dual vectors'':
\begin{eqnarray}
    V^{i}_\ell(k,\bar{x}_i) 
    &\equiv& \sqrt{ \bar{n}^i \, \frac{\Delta V_{\bar{x}_i}}{4\pi} \, P^{ii} (k) } \left[ j_\ell(k \bar{x}_i) - \beta^i j_\ell''(k\bar{x}_i) \right] \\ \nonumber
    &=& \sqrt{   \frac{\Delta V_{\bar{x}_i}}{4\pi}  \, \mathcal{P}^{ii} (k) } \left[ j_\ell(k\bar{x}_i) - \beta^i j_\ell''(k\bar{x}_i) \right] 
        \; ,
\end{eqnarray}
where we used the definition of the clustering strength introduced in Eq. \eqref{Eq:MatDef}, i.e., ${\cal{P}}^{ij}(k) = \sqrt{\bar{n}^i \bar{n}^j} \, P^{ij}(k)$. 
Using the associative property of the spectra $P^{ii} (k) P^{jj} (k')= P^{ij} (k) P^{ij} (k')$ (which is valid at least to first approximation, in the linear regime) allows us to write the data covariance as:
\begin{eqnarray}
    \label{Eq:defcovl}
    \tilde{\Gamma}^{ij}_{\ell,s}(\bar{x}_i,\bar{x}_j) 
    &=& \delta_{ij} \delta_{\bar{x}_i \, \bar{x}_j}  + 
    \sum_{[k]} V^i_\ell(k,\bar{x}_i) \, V^j_\ell(k,\bar{x}_j) \\
    \nonumber
    &=& \delta_{ij} \delta_{\bar{x}_i \, \bar{x}_j}  
    + \sum_{[k_i]}  \sum_{[k_j]}  V^i_\ell(k_i,\bar{x}_i) \, \delta^{k_i \, k_j}_F \, V^j_\ell(k_j,\bar{x}_j)
\end{eqnarray}

We also define, in analogy with the expression above, the normalized harmonic data covariance in the conjugate (Fourier) representation as: 
\begin{eqnarray}
    \label{Eq:defcovkl}
    \tilde{\Gamma}^{ij}_{\ell,s}(k_i,k_j) &=& 
    \delta_{ij} \delta^{k_i\, k_j}_F + 
    \sum_{\bar{x}} V^i_\ell(k_i,\bar{x}) \, V^j_\ell(k_j,\bar{x})
    \\ \nonumber
    &=& 
    \delta_{ij} \delta^{k_i\, k_j}_F + 
    \sum_{\bar{x}_i} \sum_{\bar{x}_j} V^i_\ell(k_i,\bar{x}_i) \,
    \delta_{\bar{x}_i \bar{x}_j}  V^j_\ell(k_j,\bar{x}_j)
    \; .
\end{eqnarray}
Just as it was in the case of real space, discussed in Section 3.2, these expressions are instrumental for the inversion of the harmonic data covariance in redshift space.
In real space the Fourier covariance is perfectly diagonal [see Eq. \eqref{Eq:Covkkpreal}], but redshift-space distortions introduce some off-diagonal structures.
However, since those off-diagonal structures in Fourier space are relatively small, the Fourier data covariance is dominated by the diagonal terms, and it becomes feasible to invert that object. All we have to do then is to transform the Fourier inverse covariance back to configuration space to obtain the inverse covariance in terms of the spherical shells.


In order to invert the expression in Eq. \eqref{Eq:defcovl} we can employ the Woodbury identity:
\begin{equation}
\begin{split}
    M &= A + XBY^{tr} \\
    \Rightarrow \quad M^{-1} &= A^{-1} - A^{-1}X(B^{-1} + Y^{tr}A^{-1}X)^{-1} Y^{tr}A^{-1} \; ,
\end{split} 
\end{equation}
Substituting $A \rightarrow \delta_{ij} \delta_{\bar{x}_i \, \bar{x}_j}, \, B \rightarrow   \delta_{ij} \delta^{k_i \, k_j}_F$ and $X,Y \rightarrow V^i_\ell (k_i, \bar{x}_i), V^j_\ell (k_j,\bar{x}_j) $,
we obtain:
\begin{eqnarray}
    [\tilde{\Gamma}^{ij}_{\ell,s}(\bar{x}_i,\bar{x}_j)]^{-1} 
    &=& \delta_{ij} \delta_{\bar{x}_i \, \bar{x}_j} \\ \nonumber
    & & -  \sum_{[k_i]}  \sum_{[k_j]} V^{i}_\ell(k_i,\bar{x}_i) 
    \left[ \delta_{ij} \delta^{k_i \, k_j}_F + \sum_{\bar{x}} V^{i}_\ell(k_i,\bar{x}) V^{j}_\ell(k_j,\bar{x}) \right]^{-1} V^j_\ell(k_j,\bar{x}_j) \; .
\end{eqnarray}
Notice that the matrix inside the square brackets is just the Fourier space harmonic covariance, $\tilde{\Gamma}^{ij}_\ell (k_i,k_j)$, defined in Eq. \eqref{Eq:defcovkl}.
Hence, if we can invert $\tilde{\Gamma}^{ij}_{\ell} (k_i,k_j)$ then we can also invert $\Gamma^{ij}_\ell (\bar{x}_i,\bar{x}_j)$:
\begin{equation}
    \label{Eq:Clx_Vk}
    [\tilde{\Gamma}^{ij}_{\ell,s}(\bar{x}_i,\bar{x}_j)]^{-1} 
    = \delta_{ij} \delta_{\bar{x}_i \, \bar{x}_j} - \sum_{ [k_i]} \sum_{[k_j]}   V^i_\ell(k_i,\bar{x}_i) \left[ \tilde{\Gamma}^{ij}_\ell(k_i,k_j) \right]^{-1} V^j_\ell(k_j,\bar{x}_j) \; .
\end{equation}

Therefore, we have reduced the problem of inverting the multi-tracer data covariance to the problem of inverting its conjugate
$\tilde{\Gamma}^{ij}_\ell(k_i,k_j)$.
Returning to Eq. \eqref{Eq:defcovkl}, we now write it in full:
\begin{eqnarray}
    \nonumber
    \tilde{\Gamma}^{ij}_\ell(k_i,k_j) & = &
    \delta_{ij}
    \delta^{k_i \, k_j}_F + 
     \sqrt{\mathcal{P}^{ii}(k_i) \, \mathcal{P}^{jj}(k_j)}
     \\ \nonumber
    & & \times \sum_{\bar{x}} \int_{\bar{x}} dx \, x^2 
    \left[  j_\ell(k_i x) - \beta^i  \, j_\ell''(k_i x) \right]
    \left[  j_\ell(k_j x) - \beta^j  \, j_\ell''(k_j x) \right]
    \\    \label{eq:approxclkk}
    &=& \delta^{k_i \, k_j}_F \left[\delta_{ij} +  (1+\beta^i) 
    (1+\beta^j) \, \sqrt{\mathcal{P}^{ii}(k_i) \, \mathcal{P}^{jj}(k_j)} \right]
    \\ \nonumber
    & & \quad + \sqrt{\mathcal{P}^{ii}(k_i) \, \mathcal{P}^{jj}(k_j)}
    \left[  \beta^i \, \beta^j  \, G_\ell(k_i,k_j) 
    - \beta^i \, H_\ell(k_i,k_j) 
    - \beta^j \, H_\ell(k_j,k_i) 
    \right] \; ,
\end{eqnarray}
where in the last line of the equations above we used the fact that $\sum_{\bar{x}} \int_{\bar{x}} dx = \int_0^\infty dx$, as well as the definitions of Eqs. \eqref{G} and \eqref{H}.
Notice that here we used again the fact that correlations (or expectation values) should be evaluated at some common {\it t} = constant hypersurface (see figure \ref{fig:LCs} and the discussion in Section 2), and the sum over radial bins in the expression above, combined with the radial integrals inside the bins, leads to an integration over the entire range of the radial coordinate ($x$) on that hypersurface, while the quantities $\bar{n}^{i}$, $b^{i}$, $\beta^{i}=b^{i}/f(z_i)$, etc., are kept fixed in that expression.
As a result, the exact same expressions $G_\ell$ and $H_\ell$ of Eqs. \eqref{G} and \eqref{H} that appeared in configuration space show up again, but now in Fourier space.
Notice also that the Fourier harmonic covariance above is an entirely different object compared with the angular power spectrum that follows from taking the expectation value of the spherical modes in a Fourier-Bessel expansion over the past light cone: in our case, the data covariance has support on the common $t=\,$ constant hypersurface where the expectation values are evaluated, not on the past light cone.

At this point it is convenient to collect the terms in the last line of Eq. \eqref{eq:approxclkk} in the Fourier (or, more accurately, {\it radial}) mixing matrix:
\begin{equation}
    \label{Eq:DefT}
    T^{ij}_\ell (k_i,k_j) \equiv \sqrt{{\cal{P}}^{ii}(k_i) {\cal{P}}^{jj}(k_j) } 
    \left[ \beta^i \beta^j \, G_\ell(k_i,k_j) 
    - \beta^i H_\ell(k_i,k_j) 
    - \beta^j H_\ell(k_j,k_i) 
    \right] \; .
\end{equation}
Eq. \eqref{eq:approxclkk} also suggests that we should define the redshift-space clustering strength:
\begin{equation}
    \mathcal{P}^{ij}_{(s)} \, \equiv \,
    (1+\beta^i)(1+\beta^j)\mathcal{P}^{ij} \; .
\end{equation}
Recall that, in the linear regime, the clustering strength can be factored, ${\cal{P}}^{ii}(k_i) {\cal{P}}^{jj} (k_j) = {\cal{P}}^{ij} (k_i) {\cal{P}}^{ij} (k_j)$, and the definition above implies that the same applies for $\mathcal{P}^{ij}_{(s)}$.
With the help of these expressions we can write finally a simple, {\it analytical} expression for the harmonic data covariance in redshift space:
\begin{eqnarray}
    \nonumber
    \tilde{\Gamma}_{\ell}^{ij}(k_i,k_j)  &=& \delta^{k_i \, k_j}_F 
    \left[ \delta_{ij} + \mathcal{P}^{ij}_{(s)} \right] 
    + T^{ij}_\ell (k_i,k_j) 
    \\ \nonumber
    & = & \delta^{k_i \, k_j}_F
    \left[ \delta_{ij} + \mathcal{P}^{ij}_{(s)}(k_i) + 
     \lambda_\ell^{ij} (k_i) \right]  
    + \tilde{T}^{ij}_\ell (k_i,k_j)
    \\ 
    \label{Eq:FourierCell}
    &=& \delta^{k_i \, k_j}_F \mathcal{S}^{ij}_\ell(k_i) 
    + \tilde{T}^{ij}_\ell (k_i,k_j) \; ,
\end{eqnarray}
where $\lambda^{ij}_\ell (k) \sim T^{ij}_\ell (k_i,k_i)$ carries the diagonal of the radial mixing matrix, and  $\tilde{T}^{ij}_\ell (k_i,k_j)$ has only off-diagonal ($k_i \neq k_j$) terms, and we defined:
\begin{equation}
    \label{Eq:defS}
    \mathcal{S}^{ij}_\ell(k) = \delta^{ij} +\lambda^{ij}_\ell +  \mathcal{P}^{ij}_{(s)} \; .
\end{equation}
Notice that both $\mathcal{P}^{ij}$ and $\lambda^{ij}_\ell$ are adimensional -- to see that this is the case for the latter, see Eq. \eqref{Eq:Discretelambda}.

\begin{figure}
    \centering
    \includegraphics[width = 0.8\textwidth]{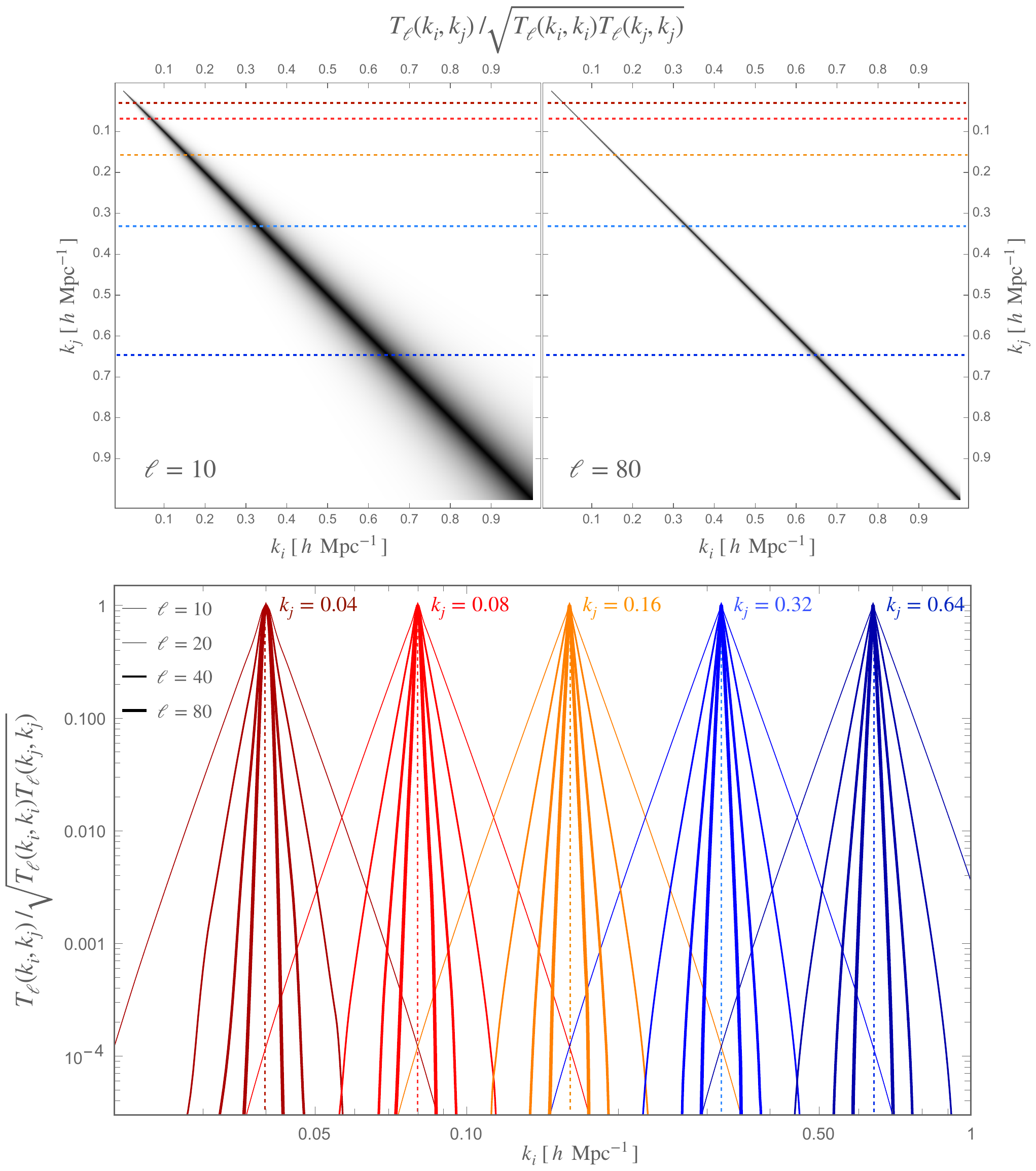}
    \caption{Radial (Fourier) mixing matrix $T_\ell(k_i,k_j)$, normalized by its diagonal for visualization purposes.
    Top: mixing matrix for $\ell=10$ (left) and $\ell=80$ (right).
    Bottom: rows of the normalized mixing matrices (denoted as the dashed lines of the two top plots), with fixed $k_j=0.04$, 0.08, 0.16, 0.32 and 0.64 $h$ Mpc$^{-1}$ (from left to right, respectively: dark red, red, orange, light blue and blue).
    Notice that for $k_i=k_j$ the normalized mixing matrix is always equal to 1.
    The different lines corresponds, from lightest to heaviest, to the multipoles $\ell=10$, 20, 40 and 80.}
    \label{fig:tmat}
\end{figure}

At this point it is useful to examine the structure of the radial mixing matrix $T_\ell^{ij}(k_i,k_j)$ defined above, in Eq. \eqref{Eq:DefT}. 
As it turns out, this matrix is, to an excellent approximation, dominated by the diagonal terms  $k_i=k_j$. 
In the top row of figure \ref{fig:tmat} we plot that matrix (normalized by the diagonal for visualization purposes) in the case of a single tracer, for the multipoles $\ell=10$ (left panel) and $\ell=80$ (right), and using $\beta=1$.
These two plots show clearly that the matrix elements decay as we move away from the diagonal.
In the lower plot of figure \ref{fig:tmat} we show the lines (of varying $k_i$) of the (normalized) radial mixing matrices with $\ell=10, 20, 40$ and $80$, for fixed values $k_j=0.04$, $0.08$, $0.16$, $0.32$ and $0.64 \, h$ Mpc$^{-1}$.
The log-scale of this bottom plot makes it clear that off-diagonal matrix elements decay as a power-law that depends on $\ell$.
It is crucial to note that the mixing matrix normalized by the diagonal, $T^{ij}_\ell (k_i,k_j)/\sqrt{T^{ii}_\ell (k_i,k_i) \, T^{jj}_\ell (k_j,k_j)}$, has no dependence at all on the shape of the power spectrum or on the number of tracers: it is determined entirely by the functions $H_\ell$ and $G_\ell$, as well as the amplitudes, $\beta^i = f(\bar{x}_i)/b^i$.
In conclusion, independently of the shape of the spectrum, and for any number of tracers, for all but the lowest multipoles it should be a very good approximation to discard the non-diagonal terms of the matrix $T_\ell^{ij}(k_i,k_j)$ -- i.e., to neglect $\tilde{T}^{ij}_\ell (k_i,k_j)$ altogether.

We can now proceed to computing the inverse of the data covariance in harmonic Fourier space given in Eq. \eqref{Eq:FourierCell}.
First, notice that:
$$
(1+\mathcal{P} + \lambda)^{-1} =
(1+\mathcal{P})^{-1} 
\left[ 1+
(1+\mathcal{P})^{-1} \lambda
\right]^{-1} \; .
$$
In a similar way to what was done in real space, in Section 3.2, using the Woodbury identity leads to the inversion:
\begin{equation}
    \label{Eq:InvPartS}
    \left[ \delta^{ij} + \mathcal{P}^{ij}_{(s)} \right]^{-1} =
    \delta^{ij} - \frac{\mathcal{P}^{ij}_{(s)}}{1+\mathcal{P}_{(s)}} \; ,
\end{equation}
where
$\mathcal{P}_{(s)}=\sum_i\mathcal{P}^{ii}_{(s)}$. 
The inverse of $\mathcal{S}^{ij}_\ell$ is therefore given by:
\begin{eqnarray}
    \label{Eq:invsk}
    [\mathcal{S}^{ij}_\ell]^{-1} &=& 
    \sum_l
    \left[ \delta^{il} - \frac{\mathcal{P}^{il}_{(s)}}{1+\mathcal{P}_{(s)}}
    \right]
    \left[ \delta^{lj} + 
    \sum_n 
    \left( \delta^{ln} - \frac{\mathcal{P}^{ln}_{(s)}}{1+\mathcal{P}_{(s)}}
    \right) \lambda^{nj}_\ell
    \right]^{-1} \; .
\end{eqnarray} 

In practical applications we would use finite Fourier bins, and it is therefore important to understand the relative importance of the redshift-space corrections $\lambda^{ij}_\ell$.
From Eq. \eqref{Eq:FourierCell} it is clear that the diagonal of the data covariance carries a Dirac delta function -- recall that in our condensed notation $\delta^{k_i k_j}_F = (\pi/2) \, \delta_D(k_i-k_j)/k_i^2$.
Using discrete Fourier bins of width $\Delta k$ we have $\delta_D (k_i-k_j) \to \delta_{k_i k_j}/\Delta k$, and
the covariance is then given by:
\begin{equation}
    \tilde{\Gamma}_{\ell}^{ij} (k_i,k_j) \to  \frac{\pi}{2}\frac{\delta_{k_i k_j}}{k_i^2\Delta k} \mathcal{S}^{ij}_\ell (k_i) + \tilde{T}^{ij}_\ell (k_i,k_j) \; ,
\end{equation}
where now we identify:
\begin{equation}
\label{Eq:Discretelambda}
\lambda^{ij}_\ell (k_i) \to \frac{2}{\pi} \, k_i^2 \, \Delta k \, T^{ij}_\ell (k_i,k_i) \; .
\end{equation}
Since the mixing matrix has dimensions of volume, the diagonal term $\lambda^{ij}_\ell$ is adimensional.
The expression above shows that these diagonal terms are proportional to the Fourier bin width, while from Eq. \eqref{Eq:DefT} and the definitions of $H_\ell$ and $G_\ell$, Eqs. \eqref{Eq:Intjj} and \eqref{eq:Hl}, it is straightforward to see that, up to factors of the redshift-space parameter $\beta$, $T_\ell^{ij} (k,k) \sim \alpha_\ell \, \mathcal{P}^{ij}(k) / k^3$,
where $\alpha_\ell$ is a coefficient that could be very large for $\ell \gg 1$. 
This means that the term $\lambda^{ij}_\ell \sim \alpha_\ell \, \mathcal{P}^{ij} (k) \, \Delta k/ k$ is in general not negligible.
If in a particular case $\lambda^{ij}_\ell$ happens to be small compared with $\delta^{ij}$ and $\mathcal{P}_{s}^{ij}$ [see Eq. \eqref{Eq:defS}], then the inverse of the matrix $\mathcal{S}^{ij}_\ell$, Eq. \eqref{Eq:invsk}, could be written in terms of a series expansion:
\begin{eqnarray}
    \label{Eq:approxclk2}
    [\mathcal{S}^{ij}_\ell]^{-1} 
    &\approx& 
    \delta^{ij} -
    \frac{\mathcal{P}^{ij}_{(s)}}{1+\mathcal{P}_{(s)}}
    -  
    \sum_{i',j'}
    \left( 
    \delta^{ii'} -
    \frac{\mathcal{P}^{ii'}_{(s)}}{1+\mathcal{P}_{(s)}}
    \right)
    \lambda^{i'j'}_\ell
    \left( 
    \delta^{j'j} -
    \frac{\mathcal{P}^{j'j}_{(s)}}{1+\mathcal{P}_{(s)}}
    \right)
    + {\cal{O}}(\lambda^2) \; .
\end{eqnarray} 
In any case, $\mathcal{S}^{ij}_\ell(k)$ is a simple $N_t \times N_t$ object (where $N_t$ is the number of tracers), which has a trivial inverse for each value of $k$ and for each multipole $\ell$.

Once the inverse $[\mathcal{S}^{ij}_\ell(k)]^{-1}$ has been determined, we can finally obtain the inverse covariance. 
Since the off-diagonal matrix $\tilde{T}^{ij}(k_i,k_j)$ is small compared with the diagonal part of the data covariance of Eq. \eqref{Eq:FourierCell}, we can invert that expression in  terms of a series expansion around $\tilde{T}^{ij}(k_i,k_j)$:
\begin{equation}
    \label{Eq:approxclk}
    [\tilde{\Gamma}_{\ell}^{ij}(k_i,k_j) ]^{-1} \approx \delta^{k_ik_j}_F 
    [\mathcal{S}^{ij}_\ell(k_i)]^{-1} - \sum_{i' , j'}
    [\mathcal{S}^{ii'}_\ell(k_i)]^{-1} \, \tilde{T}^{i'j'}_\ell (k_i,k_j) \, [\mathcal{S}_\ell^{j'j}(k_j)]^{-1}
    + {\cal{O}}(\tilde{T}^2)  \; .
\end{equation}
As discussed above, it is often sufficient to consider only the first term in the expression above, neglecting the off-diagonal matrix $\tilde{T}_\ell^{ij}$, which leads to corrections of order $\sim \, 10^{-2}-10^{-3}$, depending on the multipole.
This is a manifestation of the fact that the correlation function is much more diagonal in Fourier space than it is in real space, and is what ultimately allows us to invert the harmonic data covariance.

Before we go back to the standard notation for the data covariance, it is useful to assess the accuracy of our approximations, when we take into account the off-diagonal matrix $\tilde{T}^{ij}_\ell(k_i,k_j)$ to first order, as given in Eq. \eqref{Eq:approxclk}.
Recall that the inverse must be be such that:
\begin{eqnarray}
    \nonumber
    \sum_j \sum_{[k_j]}  \left[ \tilde{\Gamma}_{\ell}^{ij} (k_i,k_j)  \right]^{-1}
    \tilde{\Gamma}^{ji'}_{\ell}(k_j,k_{i'}) 
    &=& \sum_j
    \frac{2}{\pi} \int_0^\infty dk_j \, k_j^2 \,  \left[ \tilde{\Gamma}^{ij}_{\ell}(k_i,k_j)  \right]^{-1} \tilde{\Gamma}^{ji'}_{\ell}(k_j,k_{i'}) \\ \label{Eq:ckckinv}
    & = &\delta^{k_i \, k_{i'}}_F \delta_{ii'} 
    = \frac{\pi}{2} \frac{\delta_D(k_i-k_{i'})}{k_i^2} \delta_{ii'} \; . 
\end{eqnarray}

In order to check the accuracy of the approximation in Eq. \eqref{Eq:approxclk} we considered the case of a single tracer, such that:
$$
\mathcal{S}_\ell = 1+ \lambda_\ell + \mathcal{P}_{(s)} \quad , \quad
\mathcal{S}^{-1}_\ell = \frac{1}{1+ \lambda_\ell + \mathcal{P}_{(s)}} \; .
$$
Substituting Eq. \eqref{Eq:approxclk} into Eq. \eqref{Eq:ckckinv} then leads to a residue which is of second order in $\tilde{T}$.
We computed numerically these $\mathcal{O}(\tilde{T}^2)$ corrections in discrete Fourier bins assuming realistic scenarios for the matter power spectrum and for the redshift-space distortions, and the result is that these extra terms are negligible, typically of the order $\sim \, 10^{-6}$ compared with the lower-order terms in the expansion.
We checked the accuracy of this approximation for $\ell \lesssim 100$ and scales $k \lesssim 1 \; h \; {\rm Mpc}^{-1}$.

To summarize the results of this Section, we found that the inverse of the {\em normalized} harmonic data covariance is given by:
\begin{eqnarray}
    \label{Eq:Cinvfinal}
    \left[ \tilde{\Gamma}^{ij}_{\ell,s}(\bar{x}_i,\bar{x}_j)\right]^{-1} 
    & \simeq & \delta_{ij} \delta_{\bar{x}_i \bar{x}_j} 
    \\ \nonumber
    &-& \left( \frac{2}{\pi} \right)^2 
    \frac{\Delta V_{\bar{x}_i}}{4\pi} \frac{\Delta V_{\bar{x}_j}}{4\pi}
    \int dk_i \, k_i^2 \int dk_j \, k_j^2 \, 
    \sqrt{  \mathcal{P}^{ii} (k_i) \, \mathcal{P}^{jj}(k_j)} 
    \\ \nonumber
    & & \qquad \times \left[j_\ell(k_i\bar{x}_i) - \beta^i \, j_\ell''(k_i\bar{x}_i)\right] \left[j_\ell(k_j\bar{x}_j) - \beta^j \, j_\ell''(k_j \bar{x}_j)\right] 
    \\\nonumber
    & & \qquad \times 
    \left[
    \frac{\pi}{2} \frac{\delta(k_i-k_j)}{k_i^2} 
    [\mathcal{S}^{ij}(k_i)]^{-1}
    - \sum_{i',j'} [\mathcal{S}^{ii'}(k_i)]^{-1}
    \tilde{T}^{i'j'}_\ell (k_i,k_j)
    [\mathcal{S}^{j'j}(k_j)]^{-1} \right] \; .
\end{eqnarray}
All that is left to do is to plug the normalization factor of Eq. \eqref{Eq:NormFac} back in this expression.

\subsection{Summary of the main results}


In the previous Section we found that the inverse of the configuration-space harmonic data covariance including redshift-space distortions, Eq. \eqref{Eq:InvCovCl}, is given by: 
\begin{eqnarray}
    \label{Eq:Cinvfinal2}
    [{\Gamma}^{ij}_{\ell,s}(\bar{x}_i,\bar{x}_j)]^{-1} 
    &\simeq& 
    \frac{1}{\bar{N}^i}\,\delta_{ij} \,  \delta_{\bar{x}_i \bar{x}_j} 
    \\ \nonumber
    & &  -  \frac{2}{\pi} \int dk\,k^2 \; 
    \left[ j_\ell(k \bar{x}_i) - \beta^i j_\ell''(k\bar{x}_i) \right] 
    \left[ j_\ell(k \bar{x}_j) - \beta^j j_\ell''(k\bar{x}_j) \right] 
    \;    P^{ij}(k) \, [\mathcal{S}^{ij}(k)]^{-1}
    \\ \nonumber
    & & +  \left(\frac{2}{\pi} \right)^2 \int dk_i \,k_i^2  \int dk_j \, k_j^2 \; 
    \left[ j_\ell(k_i \bar{x}_i) - \beta^i j_\ell''(k_i \bar{x}_i) \right] 
    \left[ j_\ell(k_j \bar{x}_j) - \beta^j j_\ell''(k_j \bar{x}_j) \right] 
    \\ \nonumber
    & & \quad \times  \sqrt{P^{ii}(k_i) \, P^{jj}(k_j)} 
    \sum_{i',j'} 
     [\mathcal{S}^{ii'}(k_i)]^{-1} \tilde{T}^{i'j'}_\ell(k_i,k_j)
    [\mathcal{S}^{j'j}(k_j)]^{-1}   \; + \ldots \; ,
\end{eqnarray}
where $\mathcal{S}^{ij}(k)$ was defined in Eq. \eqref{Eq:FourierCell}, and the radial mixing matrix 
$T^{ij}(k_i,k_j)$ was defined in Eq. \eqref{Eq:DefT}.

Notice that for $\beta^i \to 0$ we also get $T_\ell^{ij} \to 0$ (so $\tilde{T}^{ij}_\ell \to 0$ and $\lambda^i_\ell \to 0$), and we recover the real-space result of Eq. \eqref{Eq:InvCovCl}.
Since $T^{ij}_\ell (k_i,k_j)$ and $S^{ij}_\ell (k)$ are given in terms of analytical expressions --- see Eqs. \eqref{Eq:DefT} and \eqref{Eq:defS} --- in a first approximation the inverse covariance above can be computed in terms of $3\times N_t(N_t+1)/2$  integrals involving Bessel functions, which can be handled either by taking the Limber approximation or numerically, using, e.g., FFTLog \cite{2000MNRAS.312..257H,fang20202d,fang2020beyond}.
If a more accurate calculation is needed, then in addition to those single integrals we must also compute $3\times N_t(N_t+1)/2$ double integrals. 
As discussed in the previous section, it is often a good approximation to neglect the matrix $\tilde{T}^{ij}_\ell$, keeping only the single Fourier integral in Eq. \eqref{Eq:Cinvfinal2}. 
Although this approximation is excellent for high multipoles ($\ell \gtrsim 10$), for very low multipoles one may need to perform the double integral in Eq. \eqref{Eq:Cinvfinal2}.

Finally, after finding at last an expression for the inverse data covariance in configuration space, including redshift-space distortions, we find ourselves in a position to go back to the Fisher matrix for the multi-tracer angular power spectra.
Since the general expressions derived in Sections 2 and 3 still hold, the Fisher matrix is given by the same expression as Eq. \eqref{Eq:fisherl}:
\begin{equation}
\label{Eq:fisherlz}
  F[C^{[ij]}_{\ell,s},C^{[i'j']}_{\ell,s}] 
  = \frac{2\ell+1}{4}
  (2-\bar\delta_{ij}) (2-\bar\delta_{i'j'})
   \left\{
   [{\Gamma}^{i i'}_{\ell,s}]^{-1}
   [{\Gamma}^{j j'}_{\ell,s}]^{-1}
   +
   [{\Gamma}^{i j'}_{\ell,s}]^{-1}
   [{\Gamma}^{i' j}_{\ell,s}]^{-1}
  \right\}   \; ,
\end{equation}
where the inverse data covariance $\left[ \Gamma^{ij}_{\ell,s} \right]^{-1}$ is given in Eq. \eqref{Eq:Cinvfinal2}.
The inverse of the Fisher matrix is formally identical to the case in real space, and leads to the multi-tracer covariance matrix for the angular power spectra in redshift space that generalizes Eq. \eqref{Eq:covl}:
\begin{equation}
\label{Eq:covlz}
  {\rm Cov}[C^{[ij]}_{\ell,s},C^{[i'j']}_{\ell,s}] 
  = \frac{1}{2\ell+1} \left[
   {\Gamma}^{i i'}_{\ell,s} \,
   {\Gamma}^{j j'}_{\ell,s}
   +
   {\Gamma}^{i j'}_{\ell,s} \,
   {\Gamma}^{i' j}_{\ell,s}
  \right]  \; .
\end{equation}
Of course, this expression can also be derived directly using the Gaussian approximation for the 4-point function of the angular power spectrum.

As discussed in this Section, the critical step is the computation of the inverse data covariances $\Gamma^{-1}$ of Eq. \eqref{Eq:fisherlz}.
But the key question is: why would one resort to these formulas, as opposed to a direct numerical computation?

Let's assume that we have 100 redshift bins and three tracers: as an example, consider a survey that maps red galaxies, blue galaxies and HI in the interval $0.5 \leq z \leq 1.5$ , in bins of width $\Delta z= 0.01$, which means comoving radius bins of widths $\Delta \chi \sim 30/E(z) \, h^-1$ Mpc, where $E(z) = H(z)/H_0$. 
That means $300 \times (299)/2 = 44850$ pairs of tracers and redshift slices.
Naively, 44850 would be the minimum number of mocks (or simulations) needed to estimate the sample covariance matrix, and depending on the nature of the tracers, a numerically stable estimation would need something of the order of $44850^2 \sim 2 \times 10^9$ mocks (see also Fig. 2, in Section 3, for an illustration of the type of matrix one would need to evaluate).
However, there may be substantial redundancy in those matrices, so perhaps a more reasonable assumption would be around $10^5$ mocks for a stable convergence of the sample covariance matrix. 
The question then is: how long would it take to produce that number of mocks/simulations (including, of course, RSDs), using the best available tools today?

The standard code to compute full-sky mocks of the past light-cone is FLASK \cite{FLASK}, or its updated version GLASS \footnote{Tessori, Loureiro \& Joachimi, in prep., \url{https://glass.readthedocs.io/en/latest/index.html}}.
We have estimated that it would take approximately $1.6 \times 10^4$ CPU hours, as well as $\sim 240$ Tb of storage, in order to produce that number of simulations with the given volume and assuming full sky, with a resolution of $Nside=512$.

On the other hand, assuming we use $\ell_{min}=2$ to $\ell_{max}=500$ (probably an overestimate, since at these radii the scales $k=\ell_{max}/\chi$ are already well into the non-linear regime), we would need to compute approximately $6 \times 5000$ double integrals in Eq. \eqref{Eq:Cinvfinal2}.
With the help of 2DFFTlog \cite{fang20202d} it is possible to compute each one of the Bessel integrals of those double integrals in approximately 2s using a single core in a desktop computer -- notice that this corresponds to all the radial bins at once for a given multipole $\ell$.
We have three of those integrals in Eq. \eqref{Eq:Cinvfinal2}, which means a total of approximately 50 hrs on a single core.

\section{Discussion and conclusions}

In this paper we present a semi-analytical expression for the multi-tracer Fisher matrix of the redshift-space angular power spectra of counts of tracers in spherical shells.
The key step that enabled us to find this expression was the fact that, according to Eq. \eqref{eqn:FishMat3}, the Fisher matrix is given by the square of the inverse {\em data} covariance, and, in the linear regime, it is possible to write an analytical expression for the inverse of the harmonic data covariance even when we include redshift-space distortions.
The main results are summarized at the end of Section 4, in Eqs. \eqref{Eq:Cinvfinal2} and \eqref{Eq:fisherlz}.

We have not resorted at any time to the Limber approximation: in fact, one of our main results (a semi-analytic expression for the inverse harmonic data covariance) follows from the fact that we were able to compute some of the radial integrals analytically.
It was because of those analytic integrals that we were able to obtain a Fourier-space radial mixing matrix that determines how the different spherical shells are correlated -- see Eq. \eqref{Eq:DefT}.
Nevertheless, our final semi-analytic expression for the inverse harmonic data covariance in redshift space, Eq. \eqref{Eq:Cinvfinal2}, would still need to be evaluated either numerically, or with the help of an approximation scheme such as the Limber approximation.

It is worth clarifying once again that our formulas are of interest mainly when a survey has a large number of thin redshift slices, and/or a large number of tracers. 
When we have only a few tracers and a small number of redshift slices, it is often more economical to simply compute the covariance of the angular power spectra, and then invert that covariance numerically.
However, with a large number of auto- and cross-spectra, the resulting high-dimensional covariance may be challenging to compute, since numerical instabilities can make it very difficult to either invert the covariance matrix, or to even allow for the computation of its Cholesky decomposition. 
Therefore, it is in the regime of many tracers and many redshift slices that our semi-analytic formulas for the Fisher matrix can become useful.

The Fisher matrix for the angular power spectra has a dual use: on one hand, it is straightforward to incorporate our results into Fisher codes in order to forecast the constraining power of future surveys, in particular those for which very large-scale phenomena are critical, such as primordial non-Gaussianities and relativistic effects. 
On the other hand, we can employ the Fisher matrix as an approximation for the uncertainties in the measurements of the angular power spectra of counts in shells, which can then be used in Markov-Chain Monte Carlo (MCMC) explorations of the likelihood in real surveys or in simulations.

A key application of our formulas, which is in fact what will ultimately justify their use in certain scenarios, is a full comparison between the constraints derived using angular power spectra, and the constraints derived using the Fourier power spectra. 
In a forthcoming paper we will compare the constraining power of the two approaches, and the biases in the parameters that either approaches may introduce.

As our final remarks, we point out that our results were derived under some key assumptions: first, we used isotropic selection functions $\bar{n}^i(\vec{x}) \to \bar{n}^i(x)$ that cover the full sky, and we assumed that the radial bins are non-overlapping, with constant tracer densities inside each radial bin, $\bar{n}^i(x) \to \bar{n}^i(\bar{x}_i)$ for all $x \in \bar{x}_i$.
Relaxing these conditions introduces couplings between the modes of the angular power spectra, and in full generality those would need to be treated using numerical methods.
However, with regards to the radial selection function, since our results apply for arbitrarily thin bins, one can always construct larger bins by summing over the smaller bins with some weights, and then taking care of the additional covariance through the appropriate Jacobian. 
And concerning a mask (or more generically, an angular selection function), which lead to coupling between the different multipoles, this can be partially mitigated by constructing bins in $\ell$. 
Another result which allows us to sustain the semi-analytical approach is the fact that the integrals of Bessel functions of different orders, which appear in the context of mode-coupling, also show a nearly diagonal behavior \cite{Maximon}.

And second, we assumed that redshift-space distortions are in the linear regime -- see Eq. \eqref{Eq:rsdxi}. 
But this is no different from what we already face when using the usual Fourier power spectrum.
Therefore, even though these are strong assumptions, this work constitute a significant first step towards simplifying the spherical description of galaxy surveys, which should further facilitate the cosmological use of the angular power spectra as a tool to study large-scale structure.

\acknowledgments
We would like to thank David Alonso, Stefano Camera and Jos\'e Fonseca for useful insights and comments on an early draft. We also acknowledge the financial support of FAPESP (R.A.), CNPq (R.A \& I.L.T.) and CAPES (J.V.D.F.). For the purpose of open access, the authors have applied a Creative Commons Attribution (CC BY) license to any Author Accepted Manuscript version arising from this submission.



\bibliographystyle{JHEP}
\bibliography{stuff.bib}

\appendix
\section{Appendix: Radial averaging of the spherical Bessel functions}

Our aim here is to explore the conditions under which the approximation of Eq.  (\ref{Eq:xijell}) can be employed. 
First of all, notice that the Fourier integration is of the form: 
$$
\int d(\ln k) j_\ell (kx_i) j_\ell (kx_j) \, k^3 P(k) \; ,
$$ 
and at low redshifts the adimensional spectrum $k^3 P(k)$ differs significantly from zero for $ k \gtrsim 10^{-2} \,  [h$ Mpc$^{-1}]$ . 
On the other hand, the comoving radii that are of most relevance for the large-scale structures are typically $x \gtrsim 10^3 \; h^{-1}$ Mpc. 
Therefore, the arguments of the spherical Bessel functions are most relevant for $kx \gtrsim 10$. 

At this point, recall the asymptotic limits of the spherical Bessel functions:
\begin{eqnarray}
    \label{Eq:jlsmall}
    \lim_{t \ll \ell} \, j_\ell(t) &=& \frac{t^\ell}{(2\ell+1)!!}
    - \frac{t^2}{2}
    \frac{t^{\ell}}{(2\ell+3)!!} + \ldots  \; , 
    \\ 
    \label{Eq:jllarge}
    \lim_{t \gg \ell} \, j_\ell(t) &=& 
    \frac{1}{t} \sin \left( t - \frac{\ell \pi}{2} \right) 
    - \frac{\ell(\ell+1)}{2t^2} \cos \left( t - \frac{\ell \pi}{2} \right) 
    + \ldots \; .
\end{eqnarray}
This means that, for $\ell \gtrsim 10$ the Bessel function varies slowly at first, according to Eq. (\ref{Eq:jlsmall}), but when $kx$ becomes of the same order as $\ell$ it starts to oscillate, eventually reaching the regime of fast oscillations given by Eq. (\ref{Eq:jllarge}). For $\ell \lesssim 10$, on the other hand, the Bessel function rapidly reaches the fast oscillating regime of Eq. (\ref{Eq:jllarge}), which further improves the convergence of the spatial integrals in Eq. (\ref{Eq:xi}).

\begin{figure}
    \centering
    \includegraphics[width = 0.48\textwidth]{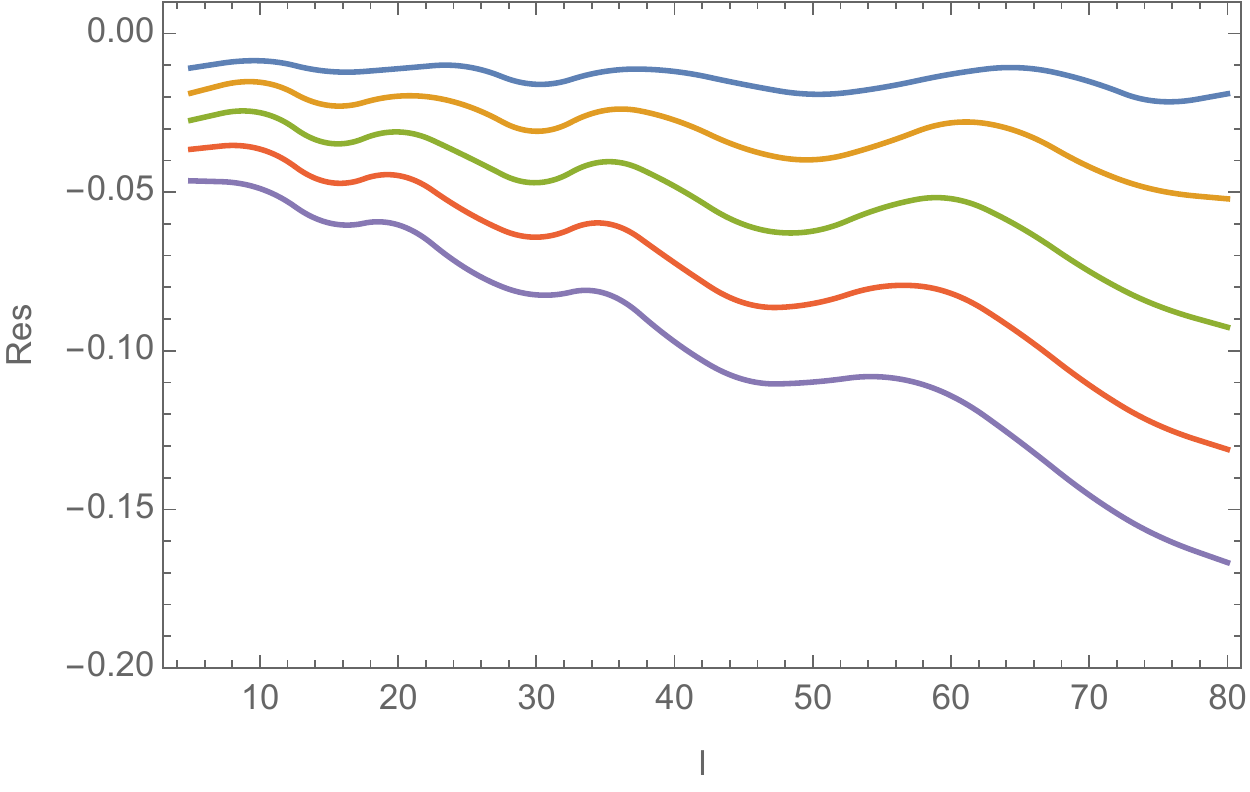}
    \includegraphics[width = 0.48\textwidth]{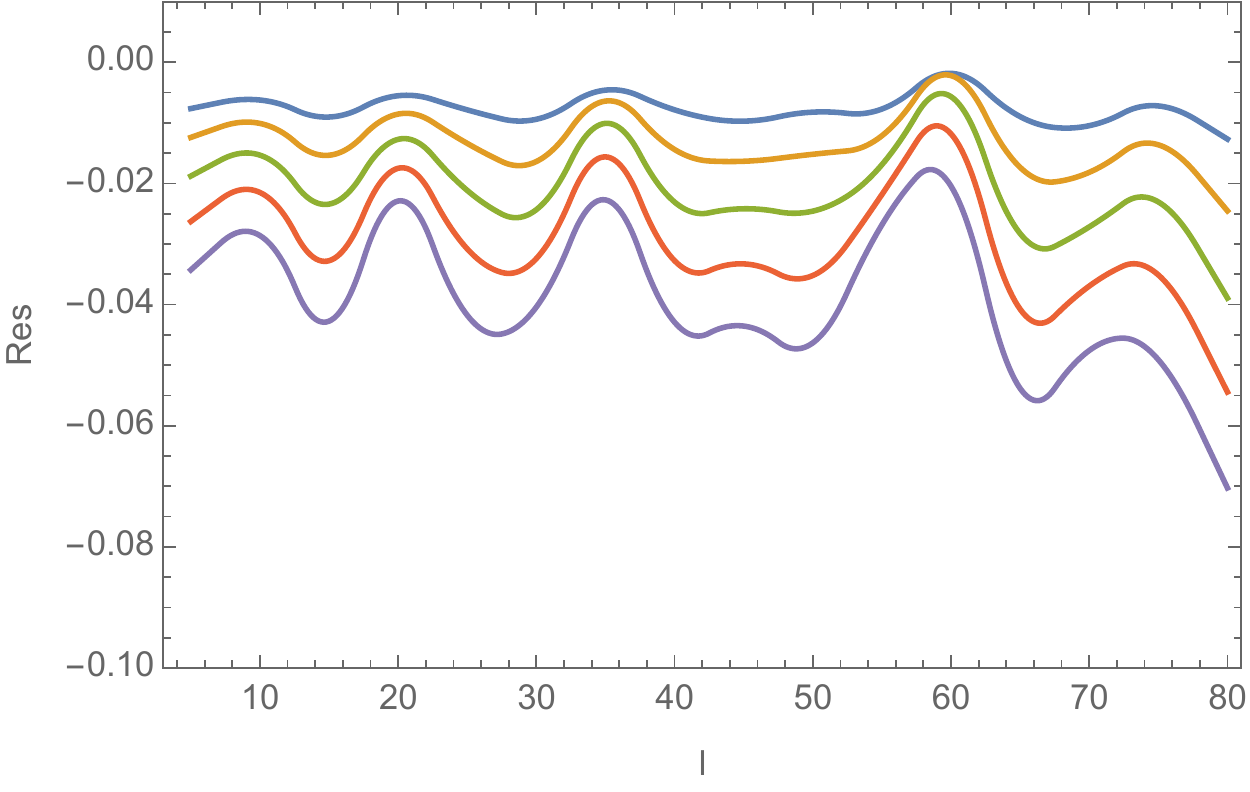}
    \caption{Residue of the approximation in Eq. (\ref{Eq:xijell}). The left and right panels show the fractional errors for $\bar{x}_i=\bar{x}_j= 10^3$ and $\bar{x}_i=\bar{x}_j= 5.
    \times 10^3 \; h^{-1}$ Mpc, respectively. The curves, from top to bottom, correspond to radial bin widths of 5, 10, 15, 20 and 25 $h^{-1}$ Mpc. For this exercise we used a $z=0$ matter power spectrum with standard cosmological parameters.}
    \label{fig:residues}
\end{figure}

In figure \ref{fig:residues} we show the accuracy of the approximation of Eq. (\ref{Eq:xijell}), in terms of the residue ${\rm Res}=C_\ell(\bar{x}_i,\bar{x}_j)/\bar{C}_\ell(\bar{x}_i,\bar{x}_j)-1$, for two values of the radii, $\bar{x}_i=\bar{x}_j= 10^3 \; h^{-1}$ Mpc (left panel) and $\bar{x}_i=\bar{x}_j=5 \times 10^3 \; h^{-1}$ Mpc (right panel), and for several radial bin widths between  $\Delta x=5$ and $\Delta x = 25 \; h^{-1}$ Mpc (for $\bar{x}_i \neq \bar{x}_j$ the amplitude of the angular power spectrum falls very fast with $|\bar{x}_i-\bar{x}_j|$, but the general trends remain the same).
The power spectrum we used for this exercise was computed using standard cosmological parameters, at $z=0$ -- notice that it is only the shape of the spectrum, not its amplitude, that matters for the accuracy of the approximation.

It is clear from figure \ref{fig:residues} that the approximation is more accurate for small values of $\ell$ as well as higher radii -- although even for the smallest bin width there is a small bias for all values of $\ell$, which comes from smearing the Fourier modes inside the radial bins.
The accuracy of Eq. (\ref{Eq:xijell}) becomes worse for progressively smaller radii, but we checked that for $\bar{x}_i=\bar{x}_j=500 \; h^{-1}$ Mpc, with $\Delta x=5 \; h^{-1}$ Mpc the approximation is accurate to within $\lesssim$5\% for $\ell < 100$. 
Conversely, these results imply that for larger radii we can indeed afford to use wider radial bins.

\section{Appendix: Toy model for the covariance and Fisher matrices}

\subsection{Analytical solutions for the angular power spectrum}

Consider the Fourier-space data covariance given, in condensed notation, by:
\begin{equation}
    \Gamma(k) = V_0 [1 + \epsilon f(k)] \; ,
\end{equation}
where $\epsilon$ is a small parameter, and $f(k)$ is any adimensional function of $k$ -- in other words, shot noise is given by $V_0$ and the power spectrum is $P(k) = \epsilon V_0 f(k)$. 
Assuming for simplicity that we have a single tracer with mean number density $\bar{n}$, and that its bias is $b=1$, the harmonic space data covariance $\Gamma_\ell(\bar{x},\bar{y})$ is given by:
\begin{eqnarray}
\nonumber
\Gamma_\ell(\bar{x},\bar{y}) 
&=& \bar{n}^2
\int_{\bar{x}} dx \, x^2
\int_{\bar{y}} dy \, y^2
\frac{2}{\pi}\int_0^\infty dk\,k^2 j_\ell(kx)j_\ell(ky) 
V_0 [1 + \epsilon f(k)]
\\ \label{eq:approxcl}
&\simeq& 
\bar{N}_{\bar{x}} \, \bar{N}_{\bar{y}} \, V_0 \, \left[
 \delta_{\bar{x} \, \bar{y}} \frac{4\pi }{\Delta V_{\bar{x}}} + 
 \epsilon \, 
\frac{2}{\pi}\int_0^\infty dk\,k^2 j_\ell(k\bar{x})j_\ell(k\bar{y}) f(k) \right]\; ,
\end{eqnarray}
where recall that $\bar{N}_{\bar{x}} = \bar{n} \, \Delta V_{\bar{x}}/4\pi$ is the number of tracers per unit solid angle in the radial bin ${\bar{x}}$, and in the passage to the second line we assumed that the widths of the radial bins are extremely small.

The inverse of the harmonic covariance is now given by:
\begin{eqnarray}
\label{eq:invcovsimple2} 
\Gamma_\ell^{-1}(\bar{x},\bar{y}) &=&  
\frac{1}{\bar{n}^2} \,
\frac{2}{\pi}\int_0^\infty dk\,k^2 j_\ell(k\bar{x})j_\ell(k\bar{y}) 
\frac{1}{V_0 [1 + \epsilon f(k)] } 
\\
\nonumber
&\simeq& 
\frac{1}{\bar{n}^2} \,
\frac{1}{V_0}
\left[
\delta_{\bar{x} \, \bar{y}} 
\frac{4\pi}{\Delta V_{\bar{x}}} - 
\, \epsilon \, 
\frac{2}{\pi}\int_0^\infty dk\,k^2 j_\ell(k\bar{x})j_\ell(k\bar{y}) f(k) + {\cal{O}}(\epsilon^2)
\right] \; .
\end{eqnarray}
Any function $f(k)$ that has an exact integral in the expression above could be used here to make an approximate calculation. E.g., a power-law such as $f(k) = 1/k$ has an exact integral. 

There is another interesting toy model that we can build using the following function:
\begin{eqnarray}
    1 + \epsilon f(k) 
    = 1 + \epsilon \frac{1}{1+ \left( \frac{k}{k_0}\right)^n}
    = \frac{1 + \epsilon + \left(\frac{k}{k_0}\right)^n }{1+ \left( \frac{k}{k_0}\right)^n} \; .
\end{eqnarray}
Notice that the inverse of the spectrum, which appears in the inverse covariance, is now basically:
\begin{eqnarray}
    \frac{1}{1 + \epsilon f(k)}
    = \frac{1+ \left( \frac{k}{k_0}\right)^n}{1 + \epsilon + \left(\frac{k}{k_0}\right)^n } 
    = \frac{\frac{1}{1+\epsilon} + \frac{1}{1+\epsilon} \left( \frac{k}{k_0}\right)^n}{1 + \frac{1}{1+\epsilon} \left(\frac{k}{k_0}\right)^n } \; .
\end{eqnarray}
Therefore, defining $1+\epsilon' \equiv 1/(1+\epsilon)$, and $k_0' \equiv k_0 (1+\epsilon)^{1/n}$, we have that the Fourier-space correlation function and its inverse are given by basically the same function:
\begin{eqnarray}
    \Gamma(k) &=& V_0 \left[  1 + \epsilon \frac{1}{1+ \left( \frac{k}{k_0}\right)^n} \right] \\
    \Gamma^{-1} (k) &=& V_0^{-1} \left[  1 + \epsilon' \frac{1}{1+ \left( \frac{k}{k_0'}\right)^n} \right] \; .
\end{eqnarray}
Therefore, we find that the covariance and the inverse covariance have terms with the same functional form:
\begin{eqnarray}
\Gamma_\ell(\bar{x},\bar{y}) 
&=& \bar{N}_{\bar{x}} \bar{N}_{\bar{y}} V_0 \left[
 \delta_{\bar{x} \, \bar{y}} \frac{4\pi }{\Delta V_{\bar{x}}} + 
 \epsilon \, 
\frac{2}{\pi}\int_0^\infty dk\,k^2 j_\ell(k\bar{x})j_\ell(k\bar{y}) \frac{1}{1+(k/k_0')^n} \right] 
\\
\Gamma_\ell^{-1}(\bar{x},\bar{y}) &\simeq& 
\frac{1}{\bar{n}^2} \,
\frac{1}{V_0}
\left[
\delta_{\bar{x} \, \bar{y}} 
\frac{4\pi}{\Delta V_{\bar{x}}} - 
\, \epsilon' \, 
\frac{2}{\pi}\int_0^\infty dk\,k^2 j_\ell(k\bar{x})j_\ell(k\bar{y}) \frac{1}{1+(k/k_0')^n} \right] \; .
\end{eqnarray}
In this way, we only need to integrate a single functional form for the two expressions. 

An interesting limit happens when we take $n\to + \infty$: in that case, we obtain a step function, and the integral becomes limited to $k \leq k_0$ for both the spectrum and its inverse (since $k_0' \to k_0$ in that limit). 
We then have:
\begin{eqnarray}
\label{eq:exactcov}
\Gamma_\ell(\bar{x},\bar{y}) 
& = & \bar{N}_{\bar{x}} \bar{N}_{\bar{y}} V_0 \left[
 \delta_{\bar{x} \, \bar{y}} \frac{4\pi }{\Delta V_{\bar{x}}} + 
 \epsilon \, 
\frac{2}{\pi}\int_0^{k_0} dk\,k^2 j_\ell(k\bar{x})j_\ell(k\bar{y}) \right]\; ,
\\
\label{eq:exactinvcov}
\Gamma_\ell^{-1}(\bar{x},\bar{y}) &=&
\frac{1}{\bar{n}^2} \,
\frac{1}{V_0}
\left[
\delta_{\bar{x} \, \bar{y}} 
\frac{4\pi}{\Delta V_{\bar{x}}} - 
\, \epsilon' \, 
\frac{2}{\pi}\int_0^{k_0} dk\,k^2 j_\ell(k\bar{x})j_\ell(k\bar{y}) \right] \; ,
\end{eqnarray}
where remember that
 $\epsilon' \equiv 1/(1+\epsilon) - 1 = -\epsilon/(1+\epsilon)$.
Now, the integral above has an exact solution \cite{bloomfield2017indefinite}:
\begin{equation}
    \int_0^1 dz \, z^2 \, j_\ell(zx) j_\ell(zy) =
    \frac{y \, j_{\ell-1} (y) \, j_{\ell} (x) - 
    x \, j_{\ell-1} (x) \, j_{\ell} (y)}{x^2-y^2} 
    \; \equiv \; g_\ell (x,y) \; ,
\end{equation}
where we defined:
\begin{eqnarray}
    g_\ell (x,y) &\equiv&
    - \frac{x\, j_{\ell-1}(x) j_\ell(y) - y\, j_{\ell-1}(y) j_\ell(x)}{x^2 - y^2} 
    \\ \nonumber
    &=& 
    \frac{ j_0(x-y) - (-1)^\ell \, j_0 (x+y) }{2\, x\,y} 
    - \frac{1}{x\,y} \sum_{n=0}^{n_\ell} [2(\ell-1-2n)+1] \, j_{\ell-2n} (x) \, j_{\ell-2n}(y) \; .
\end{eqnarray}
The sum in the second line above is over 
$n=0,1, \ldots, n_\ell = (2\ell-1)/4$, so that 
$ 2(\ell-1-2n)+1 \geq 0$.
The sum stops at $n_\ell = (\ell-1)/2$ for odd values of $\ell$, and at $n_\ell = \ell/2$ for even values of $\ell$.
This expression in the second line shows explicitly that this term is well-behaved as $x\to y$.
It is also symmetric under $x \leftrightarrow y$, since $j_0(x-y) = j_0(y-x)$ is an even function of its argument.

Therefore, we obtain that:
\begin{equation}
    \int_0^{k_0} dk \, k^2 \, j_\ell(kx) j_\ell(ky) =
    k_0^3 \, g_\ell(k_0 x,k_0y) \; .
\end{equation}
Substituting this back into Eqs. \eqref{eq:exactcov} and \eqref{eq:exactinvcov} we get:
\begin{eqnarray}
\label{eq:exactcl}
\Gamma_\ell(\bar{x},\bar{y}) 
&=& \bar{N}_{\bar{x}} \bar{N}_{\bar{y}} V_0 \left[
 \delta_{\bar{x} \, \bar{y}} \frac{4\pi }{\Delta V_{\bar{x}}} + 
 \, \epsilon \,  \frac{2}{\pi} \,
    k_0^3 \, g_\ell (k_0 \bar{x},k_0 \bar{y}) \right]\; ,
\\
\label{eq:exactclinv}
\Gamma_\ell^{-1}(\bar{x},\bar{y}) &=&
\frac{1}{\bar{n}^2} \,
\frac{1}{V_0}
\left[
\delta_{\bar{x} \, \bar{y}} 
\frac{4\pi}{\Delta V_{\bar{x}}} - 
\, \frac{\epsilon}{1+\epsilon} \, \frac{2}{\pi} \,
    k_0^3 \, g_\ell (k_0 \bar{x},k_0 \bar{y}) \right] \; .
\end{eqnarray}
It can be shown, after quite a bit of algebra, that these expressions indeed satisfy the defining condition for the inverse, Eq. (\ref{Eq:CovInvCov}).

We can in fact go one step further, and combine two "steps" like the one we used above. Let's define here the step-functions $\theta_1 = \theta_H (k-k_1)$ and $\theta_2 = \theta_H (k-k_2)$.
By ordering the k's, $k_2 > k_1$, we get that
$\theta_1^2 = \theta_1$, $\theta_2^2 = \theta_2$,  and
$\theta_1 \theta_2 = \theta_2$. 
We then define the Fourier-space correlation function as:
\begin{equation}
    \label{2.31}
    \Gamma(k) = V_0 \left( 1 + \epsilon_1 \theta_1 + \epsilon_2 \theta_2 \right) \; .
\end{equation}
 The inverse of that correlation function is now given by:
\begin{equation}
    \Gamma^{-1}(k) = V_0^{-1} \left[ 1 
    - \frac{\epsilon_1}{1+\epsilon_1} \theta_1 
    - \frac{\epsilon_2}{(1+\epsilon_1)(1+\epsilon_1+\epsilon_2)} \theta_2 
    \right] \; .
\end{equation}
It is easy to see how this expression can be used to generalize Eqs. (\ref{eq:exactcl}-\ref{eq:exactclinv}): it becomes a sum over terms like the ones we have for the case of the single step-function.

We can in fact keep adding steps to the spectrum and its inverse, and build up any solution that we want.
Using the expressions above we get:
\begin{eqnarray}
\label{eq:exactcl2}
\Gamma^\ell(\bar{x},\bar{y}) 
&=&
\bar{N}_{\bar{x}} \bar{N}_{\bar{y}} V_0 \left[
 \delta_{\bar{x} \, \bar{y}} \frac{4\pi }{\Delta V_{\bar{x}}} + 
    \, \frac{2}{\pi} \, \epsilon_1 \, 
    k_1^3 \, g_\ell(k_1 \bar{x}, k_1 \bar{y}) 
    + \,  \frac{2}{\pi} \, \epsilon_2 \, 
    k_2^3 \, g_\ell(k_2 \bar{x}, k_2 \bar{y}) 
    + \ldots
    \right] \; ,
\\
\label{eq:exactclinv2}
\Gamma_\ell^{-1} (\bar{x},\bar{y}) 
&=&
\frac{1}{\bar{n}^2} \,
\frac{1}{V_0}
\left[
\delta_{\bar{x} \, \bar{y}} 
\frac{4\pi}{\Delta V_{\bar{x}}} -  \,  \frac{2}{\pi} \, 
    \frac{\epsilon_1}{1+\epsilon_1} \, 
    k_1^3 \, g_\ell(k_1 \bar{x}, k_1 \bar{y}) \right.
    \\ \nonumber
    & & \quad \quad \quad \left.
    - \,  \frac{2}{\pi} \, \frac{\epsilon_2}{(1+\epsilon_1)(1+\epsilon_1+\epsilon_2)} \, 
    k_2^3 \, g_\ell(k_2 \bar{x}, k_2 \bar{y}) + \ldots
    \right]
 \; .
\end{eqnarray}
We can make a connection of this last equation with Eq. (\ref{eq:invcovsimple2}), where remember that $f(k) \to \bar{n} P(k)$ in that equation. We can write it as:
\begin{eqnarray}
\label{eq:invcovspec}
\Gamma_\ell^{-1} (\bar{x},\bar{y}) 
&=&  
\frac{2}{\pi}\int_0^\infty dk\,k^2 j_\ell(k\bar{x})j_\ell(k\bar{y}) 
\frac{1}{V_0} \left[ 1 - \frac{f(k)}{1 + f(k) } \right]
\\ \nonumber
&=& \frac{1}{\bar{n}^2} V_0^{-1} \left[\delta_{\bar{x} \, \bar{y}} 
\frac{4\pi}{\Delta V_{\bar{x}}}-
\frac{2}{\pi}\int_0^\infty dk\,k^2 j_\ell(k\bar{x})j_\ell(k\bar{y}) 
\frac{f(k)}{1 + f(k) } \right]
\\ \nonumber
&=& \frac{1}{\bar{n}^2} V_0^{-1} \left[\delta_{\bar{x} \, \bar{y}} 
\frac{4\pi}{\Delta V_{\bar{x}}} +
\frac{2}{\pi}\int_0^\infty dk\,k^3 g_\ell(k\bar{x},k\bar{y})
\frac{d}{dk} \left( \frac{ f}{1 + f } \right) \right]
\; ,
\end{eqnarray}
where the last equality follows from the expression in terms of the ``steps''.
So, in some sense we did an integration by parts to arrive at this last equality.

But let's go back to Eqs. (\ref{eq:exactcl2}-\ref{eq:exactclinv2}), and consider a ``top-hat'' function,  $P= \epsilon V_0 $ for $k_1 \leq k \leq k_2$, and $P=0$ outside of this interval. 
This spectral shape can be easily implemented by taking $\epsilon_1 \to \epsilon$, $\epsilon_2 \to - \epsilon$, and we obtain:
\begin{eqnarray}
\label{eq:exactcl3}
\Gamma^\ell(\bar{x},\bar{y}) 
&=&
\bar{N}_{\bar{x}} \bar{N}_{\bar{y}} V_0 \left\{ \delta_{\bar{x} \, \bar{y}} 
\frac{4\pi}{\Delta V_{\bar{x}}} + 
    \, \frac{2}{\pi} \epsilon \left[ \, 
    k_1^3 \, g_\ell(k_1 \bar{x}, k_1 \bar{y}) 
    -
    k_2^3 \, g_\ell(k_2 \bar{x}, k_2 \bar{y}) \right]
    \right\} \; ,
\\
\label{eq:exactclinv3}
\Gamma_\ell^{-1} (\bar{x},\bar{y}) 
&=&
\frac{1}{\bar{n}^2} V_0^{-1} \left\{ \delta_{\bar{x} \, \bar{y}} 
\frac{4\pi}{\Delta V_{\bar{x}}}
    - \, \frac{2}{\pi}\frac{\epsilon}{1+\epsilon} \, 
    \left[ k_1^3 \, g_\ell(k_1 \bar{x}, k_1 \bar{y}) 
    -  
    k_2^3 \, g_\ell(k_2 \bar{x}, k_2 \bar{y}) \right]
    \right\} 
 \; .
\end{eqnarray}
This is, therefore, an {\em exact} expression that can be used to compare analytic and numerical results -- and notice that $\epsilon$ does {\it not} have to be small.

\subsection{Comparing the analytical and numerical solutions}

The results of the previous section allow us to make an interesting check using the ``\textit{top-hat}'' power Spectrum of Eqs. (\ref{eq:exactcl3}-\ref{eq:exactclinv3}). 
This will allow us to compute not only the data covariance, but the covariance matrix for the angular power spectrum as well, and compare the exact result with a simulation.

In order to check the validity of our formulas we simulate Gaussian random fields in a box of $L=500$ h$^{-1}$ Mpc, with cells of $\Delta L=5$ h$^{-1}$ Mpc on the side -- i.e., we have a box with $100^3$ cells.
Our \textit{top-hat} power spectrum is given by:
\begin{equation}
    \label{Eq:analytical_ps}
    P_m(k)=\epsilon\,[\theta(k-k_1)-\theta(k-k_2)] \, ,
\end{equation}
with $\epsilon=100$ and:
\begin{equation}
    \label{Eq:analytical_ps_constants}
    \begin{gathered}
         k_1= 0.2 \, h \, {\rm Mpc}^{-1}\\
         k_2= 0.3 \, h \, {\rm Mpc}^{-1} \; .
    \end{gathered}
\end{equation}

In our simulations, for each box we draw Gaussian samples of the Fourier modes $\tilde{\delta}_m (\vec{k})$ of the density field from a probability distribution function given by the power spectrum:
\begin{equation}
    \label{Eq:gauss_pdf}
    P_m(|\tilde\delta|) \sim e^{-\frac12 \frac{|\tilde\delta_m|^2}{P(k)}}.
\end{equation}
Once the Fourier modes have been sampled in the corresponding grid, we take the inverse Fourier transform in order to obtain the matter density contrast field $\delta(\vec{r})$.
We then assume a mean density of objects in our box in order to populate each cell with a certain number of tracers, given a bias for that tracer:
\begin{equation}
    \label{Eq:tracer_density_field}
    \begin{gathered}
         N (\vec{r}) \to \bar{N} \, [1 + b \, \delta_m(\vec{r})]
         \; \longrightarrow \;
         \delta (\vec{r})=\frac{N(\vec{r})-\bar{N}}{\bar{N}} \; ,
    \end{gathered}
\end{equation}
where for simplicity we took the tracer bias $b=1$, and used $\bar{N}=10^3$ objects per cell. 
For this exercise we also used a low amplitude for the spectrum, and sigma-clipping for the very rare instances where cells end up with $1 + b \, \delta_m(\vec{r}) <0$.
The actual number of objects in each cell is finally obtained by sampling the expression above from a Poisson distribution.

Once the density field of the tracers $\delta(\vec{x})$ has been determined, we can compute its Fourier transform and take the expectation value, with the usual result that:
\begin{equation}
    \label{Eq:analytical_shotnoise_ps_general}
    \langle \tilde{\delta}(\vec{k})\,\tilde{\delta}^{*}(\vec{k}')\rangle=L^3\,\delta_{\Vec{k}\,\vec{k}'}\,
    \bigg[ P_m(k) + \frac{1}{\bar{n}} \bigg] \; .
\end{equation}

\begin{figure}
    \centering
    \includegraphics[width = 0.9\textwidth]{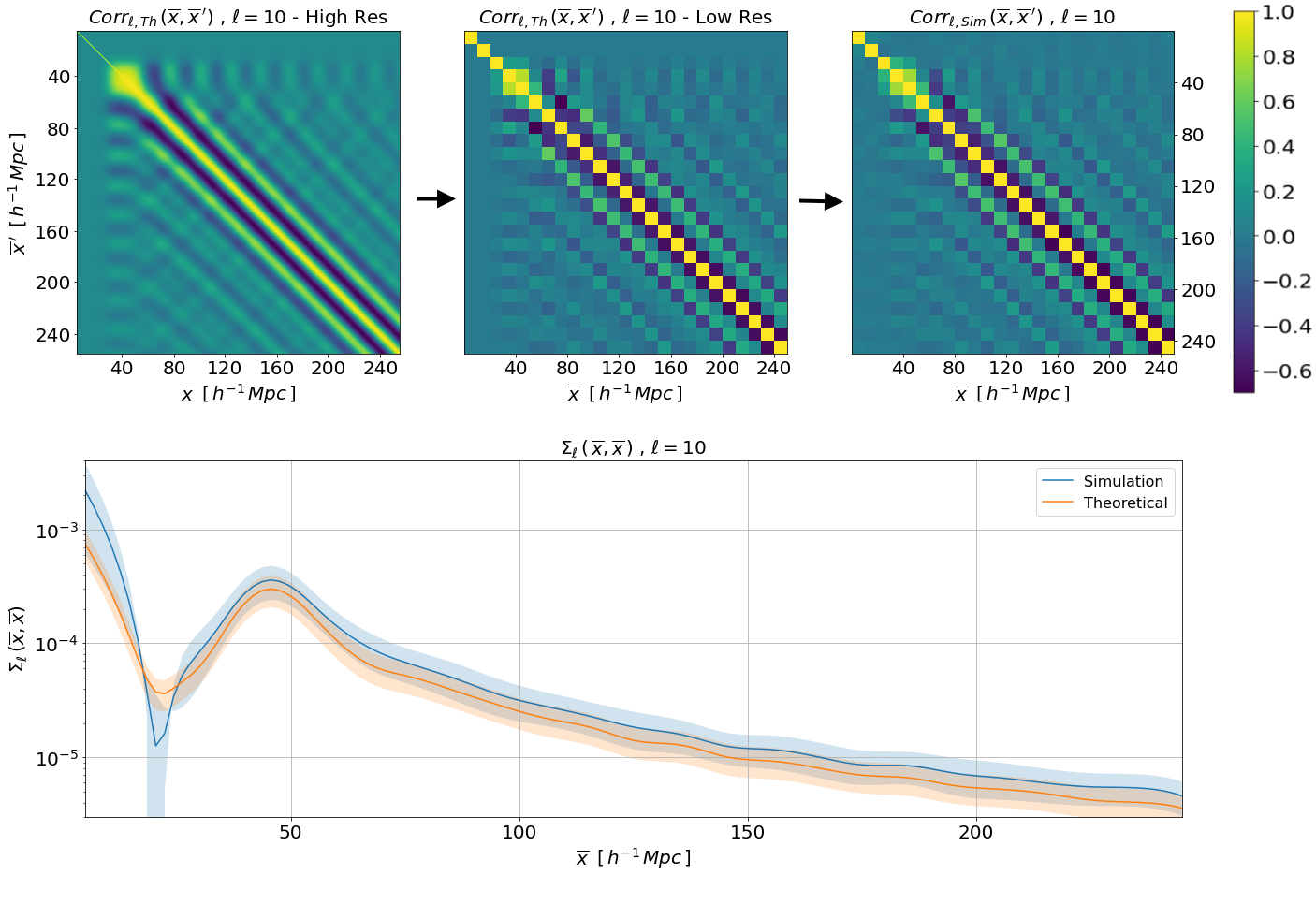}
    \caption{Harmonic data covariance  $\Gamma_\ell(\bar{x}_i,\bar{x}_j)$ derived from a top-hat Fourier power spectrum (see text), for $\ell=10$.
    The upper panels shows the harmonic data covariance $\Gamma_\ell(\bar{x}_i,\bar{x}_j)$.
    In this example we specified the number of particles such that shot noise is negligible for $\bar{x}_i \, , \bar{x}_j \gtrsim 30 \; h^{-1}$ Mpc, i.e., on large scales the data covariance is identical with the angular power spectrum $\Gamma_\ell(\bar{x}_i,\bar{x}_j) \to C_\ell(\bar{x}_i,\bar{x}_j)$.
    The plot on the top left shows a high-resolution image of the harmonic  covariance computed with the help of Eq. \eqref{Eq:sim_th_angular_ps}, normalized by its diagonal. 
    The central panel on the top row shows the result of averaging the high-resolution covariance in radial bins of $10 \; h^{-1}$ Mpc.
    The upper right panel shows the  normalized covariance in radial bins obtained from the simulations, also normalized by the diagonal.
    On the bottom panel we plot the diagonal of the two matrices. 
    The shaded regions indicate the variance around the mean: in the case of the simulation (sample mean: blue line), the sample variance was used, and in the case of the analytical result (orange line), the expression of Eq. \eqref{Eq:covl} was used. 
    The steep rise of the curves as $\bar{x} \to 0$ is due to shot noise, which falls as $\sim \bar{x}^{-2}$.
}
    \label{fig:Cl_r1r2_analytical}
\end{figure}

The next step is to compute the spherical harmonic modes $\delta_{\ell m}$ of the tracer:
\begin{equation}
    \label{Eq:analytical_harmonic_modes}
    \delta_{\ell m}(\bar{x})=\int_{\bar{x}} d^2 \hat{x} \, \delta(\vec{x})Y^{*}_{\ell m}(\hat{x})
\end{equation}
where the integral over the solid angle becomes a sum over all cells within some radial width width, $[\bar{x}-\Delta x/2,\bar{x}+\Delta x/2]$. 
In the discretization it is important to enforce the normalization by the effective solid angle subtended by the cells, $\int_{\bar{x}} d^2 \hat{x}/4\pi$.

Taking $N_s$ samples of the simulated density contrast $\delta_{\ell m}$, we can compute the tracer angular power spectrum:
\begin{equation}
    \label{Eq:sim_obs_angular_ps}
    \Gamma_{\ell,\,Sim} (\bar{x},\bar{y})
    =\dfrac{\bar{N}^2}{2\ell+1}\sum_{m=-\ell}^{\ell}\langle\delta_{\ell m}(\bar{x}) \, \delta^{*}_{\ell m}(\bar{y})\rangle_S \;,
\end{equation}
where $\langle \ldots \rangle_S$ is the sample average.

Our goal here is to compare the angular power spectrum of our simulations $C_{\ell,\,Sim}(\bar{x},\bar{y})$ to the analytical (or \textit{theoretical}) result of Eq. \eqref{eq:exactcl3}, which we will denote by $C_{\ell,\,Th}$. 
In order to draw a fair comparison, the analytical data covariance is averaged in each shell:
\begin{equation}
    \label{Eq:sim_th_angular_ps}
    C_{\ell,\,Th}(\bar{x},\bar{y})
    = 
    \bar{N}^2_{\bar{x}} \,
    \frac{\delta_{\bar{x},\bar{y}} 
    \, }{ \bar{x}^2\, \Delta x} + 
    \, \frac{2}{\pi}\,
    \epsilon \big\langle \left[ \, 
    k_1^3 \, g_\ell(k_1 x, k_1 y) 
    -
    k_2^3 \, g_\ell(k_2 x, k_2 y) \right] \big\rangle_{\bar{x},\bar{y}}
\end{equation}
where the brackets $\langle \,\ldots \rangle_{\bar{x},\bar{y}}$ denote a mean over the radial bins.

In order to compare our analytical expressions with simulations we take the multipole $\ell=10$, and use radial bins of width $\Delta x= 10 \,h^{-1} \, {\rm Mpc}$ from $\bar{x}=5 \,h^{-1} \, {\rm Mpc}$ to $\bar{x}=245\,h^{-1} \, {\rm Mpc}$. 
We produced 1000 boxes following the prescriptions described above, and computed the sample mean and sample variance of the angular power spectra. 
The comparison between the simulations and the theoretical expression is shown in figure {\ref{fig:Cl_r1r2_analytical}}: the blue and orange solid lines correspond, respectively, to the sample mean $\langle C_{\ell,Sim}(\bar{x},\bar{x}{\,}') \rangle$ and the theory data covariance from Eq. \eqref{Eq:sim_th_angular_ps} (orange).
The small difference between the simulations and the theoretical values is concentrated at small radii, where we indeed expect some smearing of the signal due to poor sampling.
For larger radii the accuracy in the shape of the angular power spectrum improves, apart from a small bias which also diminishes as we take smaller radial bins.
Moreover, the orange shaded region denotes the sample variance from the 1000 simulations, while the blue shaded region corresponds to the covariance of the angular power spectrum calculated from Eq. \eqref{Eq:covl}, where we used the 
$C_{\ell}(\bar{x},\bar{x}{\,}') $ from Eq. \eqref{Eq:sim_th_angular_ps}.

\section{Appendix: Some integrals of spherical Bessel functions}

Let's first solve the mixed integral, involving one bare and one second derivative of a spherical Bessel function, both of the same order $\ell$:
\begin{equation}
    \label{AEq:dj}
    A=\int dk\, k^2\, j_\ell(kx)j''_\ell(ky) \; .
\end{equation}
Using Eq. \eqref{ddj} we split this into three integrals:
\begin{align}\label{A}
     A&=\int\frac{dk\, k^2\, j_\ell(kx)}{(k\bar{y})^2}\bigg\{\big[\ell^2-\ell-k^2 y^2\big] j_\ell(ky)+2ky \, j_{\ell+1}(ky)\bigg\}\\
     &= \frac{(\ell^2-\ell)}{y^2}
     \underbrace{\int dk\, j_\ell(kx)j_\ell(ky)}_{A_1}
     \, - \underbrace{\int dk\, k^2 \, j_\ell(kx)j_\ell(ky)}_{A_2}\,+\,
     \frac{2}{y}
     \underbrace{\int dk\, k\, j_\ell(kx)j_{\ell+1}(ky)}_{A_3} \; .
\end{align}

For the first term, we use the following well-known result:
\begin{equation}\label{A1}
    A_1 = \int dk\, j_\ell(kx)j_\ell(ky)= \frac{\pi}{2} \frac{1}{(2\ell+1)r_>}\left( \frac{r_<}{r_>} \right)^\ell \; ,
\end{equation}
where $r_>$ and $r_<$ denote, respectively, the larger and smaller value between $x$ and $y$.
The second term is, once again, simply the closure relation:
\begin{equation}
\label{A2}
    A_2 = \int_0^\infty dk \, k^2\, j_\ell(kx)j_\ell(ky) = \frac{\pi}{2} \frac{\delta_D(x-y)}{x^2} \; .
\end{equation}    
And as for the last integral, we first express the spherical Bessel function in terms of the usual the Bessel function of the first kind:
\begin{equation}\label{jtoJ}
    j_\ell(z)=\sqrt{\frac{\pi}{2z}} J_{\ell+1/2}(z) \; ,
\end{equation}
so we can rewrite that term as:
\begin{equation}
     A_3 = \int dk\, k\, j_\ell(kx)j_{\ell+1}(ky)=\frac{\pi}{2}\,\frac{1}{\sqrt{x y}}\int dk \, J_{\ell+1/2}(kx)\,J_{\ell+3/2}(ky)  \; .
\end{equation}
This integral is also known in the literature (see, e.g., \cite{MR0010746}), and yields:
\begin{equation}
\label{A3}
    A_3 = \int dk\, k\, j_\ell(kx)j_{\ell+1}(ky)=\frac{\pi}{2}\times
\begin{dcases}
    \frac{1}{2x^2} & \text{if } x=y\\[12pt]
    0 & \text{if } x > y\\[12pt]
    \frac{x^{\ell}}{y^{\ell+2}} & \text{if } x<y \; .
\end{dcases} 
\end{equation}
Plugging these results back into Eq. \eqref{A}, we obtain:
\begin{equation}
\label{AH}
    \int dk\, k^2 j_\ell (k{x}) j_\ell''(k{y})  = -\frac{\pi}{2{x}^2}\delta_D({x}-{y}) + H_\ell({x},{y}) \; ,
\end{equation}
where:
\begin{equation}
\label{eq:AHl}
    H_\ell({x},{y}) \, =  \,
    \frac{\pi}{2y^2}
\begin{dcases}
    \frac{1}{x}\left[\frac{2(2\ell+1)+\ell(\ell-1)}{2(2\ell+1)}\right]& \text{if } {x}={y}\\[12pt]
    \frac{1}{{x}}\left(\frac{{y}}{{x}}\right)^\ell\left[\frac{2(2\ell+1)+\ell(\ell-1)}{(2\ell+1)}\right]& \text{if } {x}>{y} \\[12pt]
    \frac{1}{{y}}\frac{\ell(\ell-1)}{(2\ell+1)}\left(\frac{{x}}{{y}}\right)^\ell & \text{if } {x}<{y} \; .
\end{dcases} 
\end{equation}

The next step is to find an analytical expression for the integral which contains two second derivatives of spherical Bessel functions:
\begin{equation}
\nonumber
\int dk\, k^2 j_\ell''(k {x})j_\ell''(k {y}) \; .
\end{equation}
In similar fashion to what we did for Eq.  \eqref{AEq:dj}, we rewrite the derivatives and split this integral into nine parts:
\begin{equation}
\nonumber
\begin{gathered}
     \int\frac{dk}{k^2 {x}^2 {y}^2}
     \left\{\left[\ell^2-\ell-k^2 {y}^2\right]\, j_\ell(k {y})+2k {y}\, j_{\ell+1}(k {y})\right\}
     \left\{\left[\ell^2-\ell-k^2 {x}^2\right]\, j_\ell(k {x})+2k {x}\, j_{\ell+1}(k {x})\right\}\\
     = 
     \frac{(\ell^2-\ell)^2}{( {x} {y})^2}
     \underbrace{\int \frac{dk}{k^2}\, j_\ell(k {x})j_\ell(k {y})}_{B_1}\,-\,
     \frac{(\ell^2-\ell)}{ {y}^2}
     \underbrace{\int dk\, j_\ell(k {x})j_\ell(k {y})}_{B_2}\,+\,
     2\frac{(\ell^2-\ell)}{ {x} {y}^2} 
     \underbrace{\int \frac{dk}{k} j_{\ell+1}(k {x})j_\ell(k {y})}_{B_3}
     \\
     \,\,-\frac{(\ell^2-\ell)}{ {x}^2}
     \underbrace{\int dk\,j_\ell(k {x})j_\ell(k {y})}_{B_4}\quad+\quad
     \underbrace{\int dk\,k^2\, j_\ell(k {x})j_\ell(k {y})}_{B_5}
     \quad-\quad
     \frac{2}{ {x}}
     \underbrace{\int dk\,k\, j_{\ell+1}(k {x})j_\ell(k {y})}_{B_6}\\
     \quad+
     2\frac{(\ell^2-\ell)}{ {y} {x}^2} 
     \underbrace{\int \frac{dk}{k} j_\ell(k {x}) j_{\ell+1}(k {y})}_{B_7}\,-\,
     \frac{2}{ {y}}
     \underbrace{\int dk\,k\,j_\ell(k {x})  j_{\ell+1}(k {y})}_{B_8}\,+\,
     \frac{4}{ {x} {y}}
     \underbrace{\int dk\, j_{\ell+1}(k {x})j_{\ell+1}(k {y})}_{B_9}
\end{gathered}
\end{equation}

Most of the integrals above are known, except for $B_1$ and $B_3$ (which is identical to $B_7$). Although those two integrals have analytical formulas in terms of hypergeometric functions \cite{MR0010746}, we have found more suitable expressions for these integrals when $\ell$ assumes integer values. 
We start by rewriting the integral in $B_1$ as:
\begin{equation}
    B_1 = r_>\int_0^\infty \frac{dq}{q^2}\, j_\ell(q)j_\ell\left(q\frac{r_<}{r_>}\right)
\end{equation}
with $q = kr_>$, and after many manipulations we find the expression:
\begin{equation}\label{B1}
    B_1 = \frac{\pi}{2} \frac{2^\ell(\ell+1)!(2\ell-3)!!}{(2\ell+3)!}r_>\left(\frac{r_<}{r_>} \right)^\ell \left[2\ell+3-(2\ell-1)\left( \frac{r_<}{r_>} \right)^2 \right] \; .
\end{equation}
Similarly, for the integral $B_3$ after some algebra we obtain the analytical expression:
\begin{equation}
\label{B3}
    B_3 = \frac{\pi}{2} \; 
\begin{dcases}
    \frac{1}{(2\ell+1)(2\ell+3)}& \text{if }  {x}= {y}\\[12pt]
    \frac{1}{(2\ell+1)(2\ell+3)}\left(\frac{ {x}}{ {y}}\right)^{\ell+1} & \text{if }  {x}< {y} \; ,
    \\[12pt]
    \frac{1}{2}\left( \frac{ {y}}{ {x}} \right)^{\ell}\left[ \frac{1}{2\ell+1} - \frac{1}{2\ell+3}\left( \frac{ {y}}{ {x}} \right)^2 \right] & \text{if }  {x}> {y} \; .
\end{dcases} 
\end{equation}
We checked these expressions against numerical integrations with the help of FFTLog \cite{2000MNRAS.312..257H, fang20202d,fang2020beyond}, from $\ell=2$ up to $\ell=50$, and verified that they are correct.
We were unable to ascertain whether these closed forms of the integrals $B_1$ and $B_3$ were previously known.
   
Collecting these results, together with Eqs. \eqref{A1}, \eqref{A2} and \eqref{A3}, we can finally write an analytical expression for the integral:
\begin{equation}\label{AG}
    \int_0^\infty dk \, k^2 \, j''_\ell(k {x})j''_\ell(k {y})
    = \frac{\pi}{2} \frac{\delta_D( {x}- {y})}{ {x}^2} + G_{\ell}( {x}, {y}) \; ,
\end{equation}
where:
\begin{eqnarray}
    \label{Eq:AIntjj}
    G_{\ell}( {x}, {y})
    & = & 
    \frac{\pi}{2} \frac{r_<^\ell}{r_>^{\ell+1}} 
    \left\{ 
    \left[ \frac{4}{(2\ell+3)} -2 \right] \frac{1}{r_>^2} \right.
    \\     \nonumber
    & & + \frac{2 \, \ell(\ell-1)}{(2\ell+1)(2\ell+3)} 
    \frac{1}{r_>^2}    - 
    \frac{\ell(\ell-1)}{(2\ell+1)}
    \left( \frac{1}{r_>^2} + \frac{1}{r_<^2} \right)
    \\
    \nonumber
    & & + \frac{2 \, \ell(\ell-1)}{r^2_< }  \left[ \frac{1}{(2\ell+1)} - \frac{1}{(2\ell+3)} \left( \frac{r_<}{r_>} \right)^2  \right] \\ 
    \nonumber
    & & \left.
    + 2^\ell [\ell(\ell-1)]^2 
     \frac{(\ell+1)!(2\ell-3)!!}{(2\ell+3)!}
     \frac{1}{r_< ^2}
     \left[ 2\ell+3-(2\ell-1)\left( \frac{r_<}{r_>} \right)^2 \right]  \right\} \; .
\end{eqnarray}


\end{document}